\PassOptionsToPackage{dvipsnames}{xcolor} 
\documentclass[twocolumn]{aastex631}

\usepackage{graphicx}	          
\usepackage{amsmath}	          
\usepackage{upgreek}            
\usepackage{amssymb}	          
\usepackage{bm}		              
\usepackage{CJKutf8}            
\usepackage[T1]{fontenc}        
\usepackage{comment}            
\usepackage{tabularx}           
\usepackage{threeparttable}     
\usepackage{booktabs}           
\usepackage{soul}               
\usepackage{ulem}               
\usepackage{multirow}           
\usepackage{makecell}
\usepackage{microtype}          
\DisableLigatures[f]{encoding = *, family = *} 
\usepackage[figuresleft]{rotating}  
\allowdisplaybreaks             
\emergencystretch=\maxdimen     
\usepackage{etoolbox}
\makeatletter
\patchcmd{\ltx@foottext}{%
  .5\textwidth\advance\hsize-18pt}{%
  \linewidth\advance\hsize-1.8em%
}{}{}
\makeatother

\def\numx#1e#2{{#1}\mathrm{e}{#2}}

\newcommand{\rlr}[1]{#1}

\defcitealias{Dai2023}{D23}

\received{--}
\revised{--}
\accepted{--}

\shorttitle{Resonant Remains of Broken Chains}
\shortauthors{Li et al.
}

\begin{document}

\title{The Resonant Remains of Broken Chains from Major and Minor Mergers}

\correspondingauthor{Rixin Li}
\email{rixin@berkeley.edu}

\author[0000-0001-9222-4367]{Rixin Li
\begin{CJK*}{UTF8}{gkai}(李日新)\end{CJK*}}
\altaffiliation{51 Pegasi b Fellow}
\affiliation{Department of Astronomy, Theoretical Astrophysics Center, and Center for Integrative Planetary Science, University of California Berkeley, Berkeley, CA 94720-3411, USA}

\author[0000-0002-6246-2310]{Eugene Chiang
\begin{CJK*}{UTF8}{bkai}(蔣詒曾)\end{CJK*}}
\affiliation{Department of Astronomy, Theoretical Astrophysics Center, and Center for Integrative Planetary Science, University of California Berkeley, Berkeley, CA 94720-3411, USA}
\affiliation{Department of Earth and Planetary Science, University of California Berkeley, Berkeley, CA 94720-4767, USA}

\author[0000-0003-0690-1056]{Nick Choksi}
\affiliation{Department of Astronomy, Theoretical Astrophysics Center, and Center for Integrative Planetary Science, University of California Berkeley, Berkeley, CA 94720-3411, USA}

\author[0000-0002-8958-0683]{Fei Dai
\begin{CJK*}{UTF8}{gkai}(戴飞)\end{CJK*}}
\affiliation{Institute for Astronomy, University of Hawai`i, 2680 Woodlawn Drive, Honolulu, HI 96822, USA}

\begin{abstract}
{\it TESS} and {\it Kepler} have revealed that practically all close-in sub-Neptunes form in mean-motion resonant chains, most of which unravel on timescales of 100 Myr.  Using $N$-body integrations, we study how planetary collisions from destabilized resonant chains produce the orbital period distribution  observed among mature systems, focusing on the resonant fine structures remaining post-instability.  In their natal chains, planets near first-order resonances have period ratios just wide of perfect commensurability, driven there by disk migration and eccentricity damping.  Sufficiently large resonant libration amplitudes are needed to trigger instability.  Ensuing collisions between planets (``major mergers'') erode but do not eliminate resonant pairs; surviving pairs show up as narrow ``peaks'' just wide of commensurability in the histogram of neighboring-planet period ratios.  Merger products exhibit a broad range of period ratios, filling the space between relatively closely-separated resonances such as the 5:4, 4:3, and 3:2, but failing to bridge the wider gap between the 3:2 and 2:1 --- a ``trough'' thus manifests just short of the 2:1 resonance, as observed.  Major mergers  generate debris which undergoes ``minor mergers'' with planets, in many cases further widening resonant pairs.  With all this dynamical activity, free eccentricities of resonant pairs, and by extension the phases of their transit timing variations, are readily excited.  Non-resonant planets, being merger products, are predicted to have higher masses than resonant planets, as observed.  At the same time, a small fraction of mergers produce a high-mass tail in the resonant population, also observed.
\end{abstract}

\section{Introduction}
\label{sec:intro}

\begin{figure*}
  \centering
  \includegraphics[width=\linewidth]{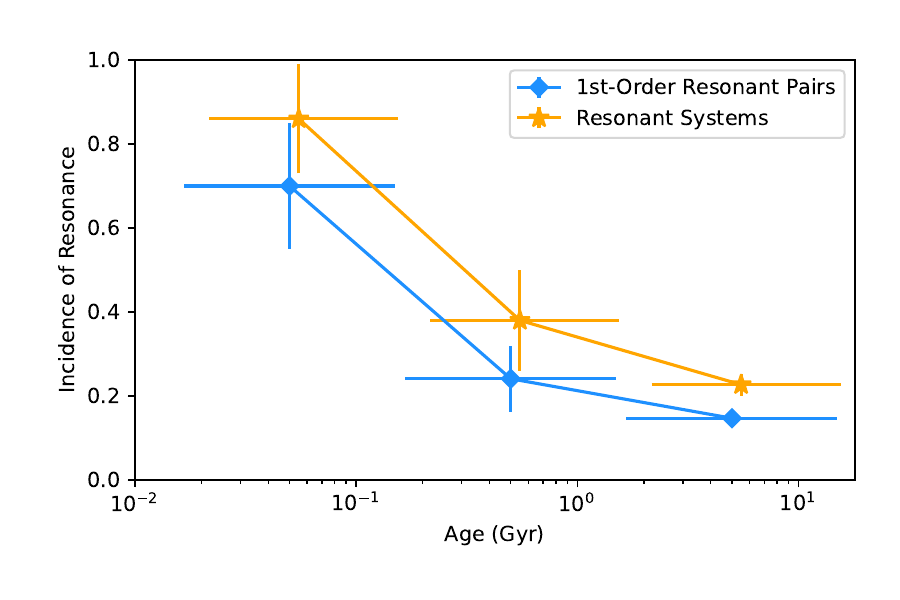}
  \caption{Reprinted from Figure 7 of \citet{Dai2024}: the observed decline with age of the
    fraction of neighboring planet pairs in 1st-order resonance (blue diamonds), and the concomitant decline in the fraction of planetary
    systems containing at least one pair in a 1st or 2nd-order resonance (orange
    stars, offset from blue diamonds for clarity). \citet{Dai2024} classify a pair as being in 1st-order
    resonance if $-0.015 \leq  \Delta \leq 0.03$, and in 2nd-order
    resonance if $-0.015 \leq \Delta \leq 0.015$.
  \label{fig:resonance_fraction}}
\end{figure*}

By detecting planetary systems in young stellar clusters, the {\it Transiting Exoplanet Survey Satellite} \citep[{\it TESS}; ][]{Ricker2014} has effectively enabled us to travel back in time, to see that most close-in exoplanets, with orbital periods $\lesssim 100$ days, formed in mean-motion resonance.  Figure \ref{fig:resonance_fraction}, reproduced from \citet{Dai2024}, shows that the fraction of neighboring exoplanet pairs that lie near 1st-order resonance, with period ratios within a few percent of $(m+1)$:$m$ for positive integer $m$, is as high as $\sim$70\% at ages between $10^7$--$10^8$ yr.  The fraction of systems containing at least one pair in 1st or 2nd-order [$(m+2)$:$m$] resonance is still higher, $\sim$86\%.  These fractions decline to $\sim$15-23\% at ages $> 10^9$ yr \citep[see also][]{Lissauer2011, Fabrycky2014}.

Resonances are a smoking gun for migration; it seems clear that close-in exoplanets, most of which are Neptune-sized or smaller, underwent convergent migration in their parent disks and captured themselves into mean-motion resonance, on timescales $\lesssim 10^7$ yr.  The occasional resonant pairs that we observe around Gyr-old stars, and the handful of similarly old and spectacularly long resonant chains
(e.g.~Kepler-60, \citealt{Jontof-Hutter2016};
Kepler-80, \citealt{MacDonald2016, Weisserman2023};
Kepler-223, \citealt{Mills2016};
TRAPPIST-1, \citealt{Agol2021, Teyssandier2022};
TOI-178, \citealt{Leleu2021};
and TOI-1136, \citealt{Dai2023}), are understood to be the survivors of dynamical instabilities that broke apart primordially long chains on timescales of 100 Myr \citep[e.g.][]{Izidoro2017, Izidoro2021, Goldberg_Batygin_2022,Liveoak2024}.

\begin{figure*}
  \centering
  \includegraphics[width=\linewidth]{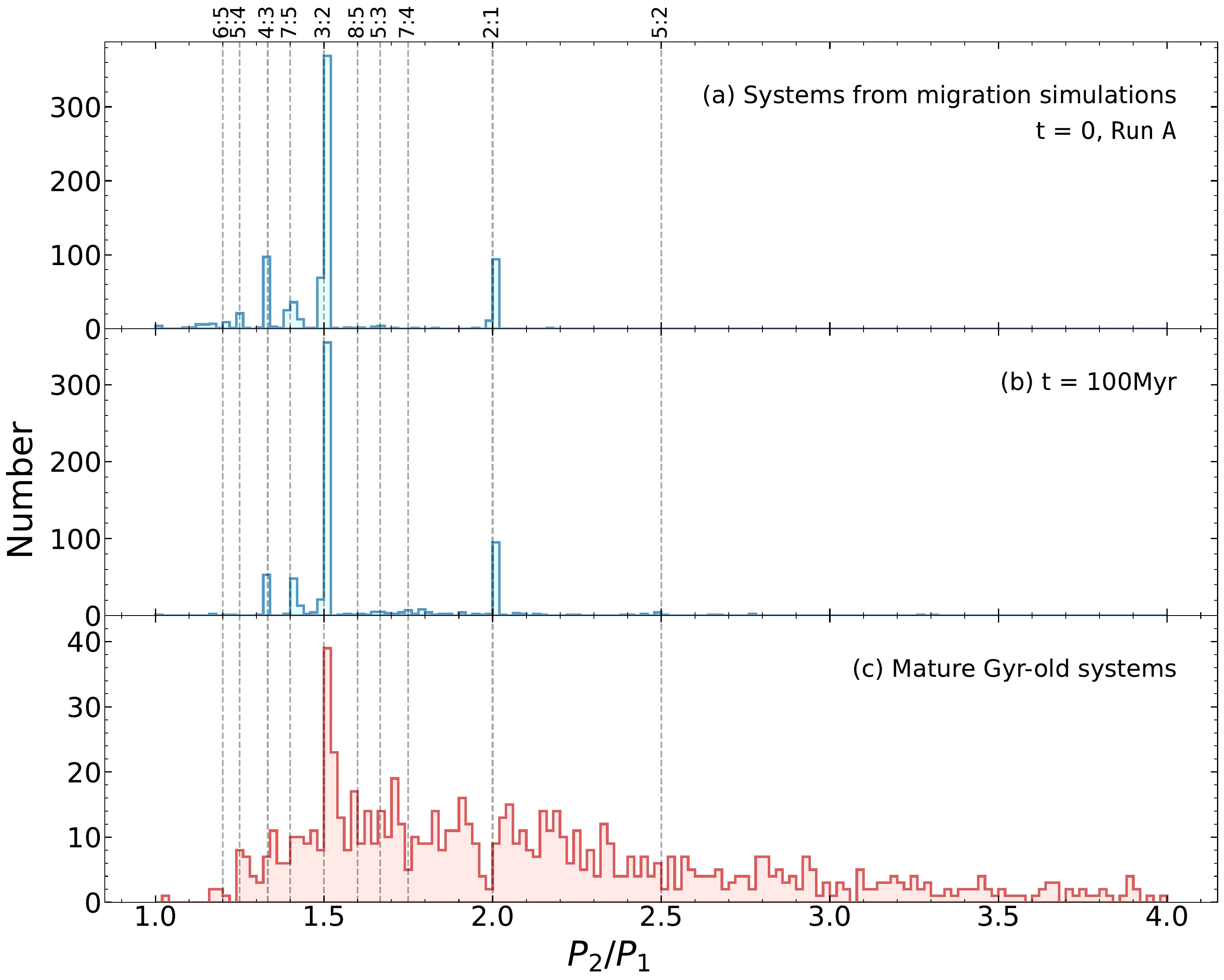}
  \caption{Period ratios $P_2/P_1$ of neighboring planets, either observed today for planets with radii $< 4R_\oplus$ in the March 2024 NASA Exoplanet Archive (bottom panel), or simulated from \texttt{Run A}, which is initialized with TOI-1136 analogues from the disk-planet simulations of \citetalias{Dai2023} (top panel), and integrated for 100 Myr (middle panel).  The \texttt{Run A} systems mostly do not change over time; they are too tightly locked in resonance.  The histogram bins have widths of 0.02.
  \label{fig:mig_sims}}
\end{figure*}

Our present study seeks to better understand how resonant chains break to give rise to the properties of mature planetary systems, considered en masse.  We do not study the important question of why they break (i.e.~the physical causes of instability), but concentrate instead on what happens after they break (i.e. the consequences of instability).  We focus on reproducing two observables: the distribution of orbital period ratios $P_2/P_1$ of adjacent planets (where $P_1$ is the orbital period of the inner member of a pair, and $P_2$ the period of the outer member), and the transit timing variations of near-resonant pairs.\footnote{In this paper, unless otherwise indicated, we use the terms ``resonant'' and ``near-resonant'', and ``resonance'' and ``commensurability'', interchangeably.  Often we care only that a pair has a period ratio close to (within a few percent of) an $(m+1)$:$m$ or $(m+2)$:$m$ commensurability, and not necessarily whether its resonant argument is librating or circulating.} The former is displayed in the bottom panel of Figure \ref{fig:mig_sims}, plotted for planets with radii $< 4 R_\oplus$ orbiting Gyr-old stars (\citealt{PSCompPars}).  The salient features of the $P_2/P_1$ period ratio histogram are: (i) a broad continuum of values atop of which are (ii) excess populations near 1st-order commensurabilities, 3:2 being the most prominent, followed by the 2:1, 4:3, and 5:4, all of which have (iii) a preference for period ratios slightly wide of perfect commensurability.  For positive integers $p$ and $q$, the fractional deviation from a perfect $p$:$q$ period commensurability is measured by
\begin{equation}
  \Delta \equiv \frac{q}{p} \frac{P_2}{P_1} - 1 \,.
\end{equation}
The resonant ``peaks'' in the histogram have $\Delta \sim +1\%$ (histogram bins in Figure \ref{fig:mig_sims} have widths of 2\%).  The 2:1 peak---and notably not the other peaks at 3:2, 4:3, and 5:4---is accompanied by a ``trough'' (a population deficit) at $\Delta \sim -1\%$.  Peak-trough structures can be formed from dissipative processes like stellar tides \citep{Lithwick_Wu_2012, Batygin_Morbidelli_2013} and dynamical friction exerted by parent disks \citep[e.g.][]{Choksi2020}--- any process that drives pairs to their resonant fixed points, which naturally lie at $\Delta > 0$ (because resonant orbits apsidally regress; see the introduction of \citealt{Choksi2020}).  We will examine whether these structures in the $P_2/P_1$ histogram can survive, or be created by, the dynamical upheavals that broke resonant chains.

For the other observable --- transit timing variations (TTVs) in mature resonant pairs --- we restrict our attention to TTVs that vary sinusoidally, which requires that pairs be sufficiently isolated from other mean-motion resonances with other planets.  For sinusoidal TTVs, ``phases'' can be evaluated (\citealt{Lithwick_Xie_Wu_2012, Choksi2023}).  The TTV phase measures ``free'' eccentricities, i.e., the components of the eccentricity vector that are specified by initial conditions and not by the mutual forcing between bodies (by analogy to the free and forced components of a driven harmonic oscillator).  The larger the free eccentricities, the more the resonant argument varies with time; the pair may be in large-amplitude libration or circulation.  Eccentricities and periapse longitudes, and by extension resonant arguments, are difficult to measure from transit measurements alone, and thus the TTV phase (the measurement of which we will describe in section \ref{sec:ttv_phases}) serves as a convenient proxy for free eccentricity and libration/circulation.  The observational puzzle is why observed TTV phases are large ($\gtrsim$ 1 rad in magnitude; \citealt{Lithwick_Xie_Wu_2012}; \citealt{Hadden_Lithwick_2014}; \citealt{Choksi2023}), when dissipative processes like disk-driven migration and eccentricity damping --- which otherwise adequately explain peak-trough structures in the $P_2/P_1$ histogram \citep{Lithwick_Wu_2012, Batygin_Morbidelli_2013, Choksi2020} --- would damp away free eccentricities and by extension TTV phases.  The dynamical instabilities that dissolved resonant chains are obvious suspects for exciting TTV phases, and we will bring them in for questioning.

Our main interrogation tool is $N$-body integrations.  These are staged after planets form and assemble into resonant chains, last 100 Myr in disk-free environments (we would integrate for longer if not for the computational cost), and involve hundreds to thousands of variously initialized multi-planet systems to statistically survey the chaotic evolution in phase space.  We do not model the earlier phase of planet formation in disks \citep[cf.][]{Izidoro2017, Izidoro2021}, nor do we attempt to dissect mechanisms for instability \citep[cf.][]{Goldberg_etal_2022}.  Instead we experiment with different post-disk initial conditions to try to identify those needed to generate instability and reproduce the observations.  Instability and orbit crossing lead, for sub-Neptunes this deep in the star's potential well, to planetary collisions.  We follow the outcomes of mergers, and experiment with a rudimentary treatment of planetesimal debris created from mergers, inspired by \citet{Wu2024} who point out how peak-trough structures and TTV phases may arise from scattering of debris.  The evolutionary era we are attempting to simulate is akin to that of the ``giant impacts'' phase in our Solar System \citep[see the review by][]{Gabriel_Cambioni_2023}.  Our work is in the same vein as \citet{Goldberg_Batygin_2022}, who ask whether instabilities and the mergers resulting therefrom can explain the claimed intra-system uniformity of {\it Kepler} planets.  These authors also examined the observed distribution of period ratios; our analysis will be more pointed as we focus on the finer structures in the period histogram, in particular the 2:1 peak-trough asymmetry.

In many (but not all) of our experiments we adopt as our template for initial conditions the real-life resonant chain TOI-1136 \citep{Dai2023}.  This six-planet chain is attractive for several reasons.  It features the most populous resonances observed among (mature) {\it Kepler} planets, the 3:2 and 2:1, in about the right proportions (three 3:2 vs.~one 2:1).  It is $\sim$700 Myr old, an order of magnitude younger than other resonant chains like TRAPPIST-1 and TOI-178, and its properties are therefore arguably more reflective of formation conditions.  Also, unlike other chains, it enjoys the most complete, publically available set of vetted orbital elements (including mean anomalies).  In this last regard, the orbital eccentricities of TOI-1136 are several times larger than those of other resonant chains, which promotes instability (as we have verified by direct experimentation).  We will leverage the work of \citet{Dai2023}, who provide a number of disk-planet simulations designed to reproduce TOI-1136 (their section 6.3), plus a convenient set of possible ranges for its orbital parameters (their Table 10).

In using TOI-1136 as a basis for our long-term integrations, we are assuming a kind of ergodic hypothesis, that an individual system (or in our case, slightly different variants of an individual system) can, over time, access the same phase space available to a diverse ensemble of systems.  We will see the limitations of this hypothesis, motivating an additional set of experiments that relax the assumption of a TOI-1136 architecture to more fully randomize initial conditions.

This paper is organized as follows.  We begin in section \ref{sec:int_mig_sim} with an {\it a priori} approach, initializing our $N$-body integrations with the results of disk-migration and eccentricity damping simulations from \citet{Dai2023} that reproduce TOI-1136 to varying degrees.  From these integrations we will see whether we can reproduce the observed decline in resonance occupation (Fig.~\ref{fig:resonance_fraction}), and the various observed features of the $P_2$/$P_1$ period ratio histogram for mature planets (bottom panel of Fig.~\ref{fig:mig_sims}).  In section \ref{sec:int_empirical}, we try a semi-empirical approach, basing our $N$-body integrations directly on observed orbital elements of TOI-1136, and in particular their ranges as allowed by observational uncertainty.  This approach will prove more fruitful, and we will experiment with further tweaks in initial conditions, and add ``minor mergers'' with planetesimal debris generated from ``major mergers'' between planets, in the effort to better reproduce the observations.  Section \ref{sec:random_experiment} replaces the TOI-1136 template with a much more diverse set of initial conditions in a final attempt to reproduce all the features of the $P_2/P_1$ histogram.  Section \ref{sec:ttv_phases} compares predicted TTV phases with those observed.  We conclude in section \ref{sec:summary}.

\begingroup
\setlength{\medmuskip}{0mu}
\begin{deluxetable*}{clr}
  \tabletypesize{\normalsize}
  \tablecaption{100-Myr Integrations}\label{tab:runs}
  \tablecolumns{3}
  \tablehead{
    \colhead{\texttt{\bf Run}} &
    \colhead{Setup} &
    \colhead{N}
  }
  \startdata
  \hline
   \texttt{\bf A} & Analogues of TOI-1136 from migration and eccentricity damping simulations in \citetalias{Dai2023}, & 160 \\
                  & first prescription of their section 6.3 & \\
   \texttt{\bf B} & Analogues of TOI-1136 from best-fit observations and uncertainty ranges in \citetalias{Dai2023}, & 256 \\
                  & using $\Delta_+$ in Table \ref{tab:toi-1136} & \\
   \texttt{\bf B1} & Similar to \texttt{\bf B} but replacing e:f=7:5 with e:f=3:2 & 256 \\
   \texttt{\bf B2} & Similar to \texttt{\bf B} but replacing e:f=7:5 with e:f=2:1 & 256 \\
   \texttt{\bf B3} & Similar to \texttt{\bf B} but with initial eccentricities damped to $<0.01$ & 256 \\
   \texttt{\bf C} & Similar to \texttt{\bf B} but with $N_{\rm debris} = 30$ coplanar debris particles generated per major merger & 256 \\
   \texttt{\bf D} & Similar to \texttt{\bf C} but with debris particles isotropically ejected in 3D & 256 \\
   \texttt{\bf D1} & Similar to \texttt{\bf D} but with a Rayleigh distribution of initial planet inclinations ($\sigma = 0.5^\circ$) & 256 \\
   \texttt{\bf E} & A $\chi^2$-selected ensemble of initially non-coplanar systems with diverse resonant architectures & 294 \\
   \rlr{\texttt{\bf E1}} & \rlr{Similar to \texttt{\bf E} but with no $\chi^2$ selection} & \rlr{317} \\
  \hline
  \enddata
\end{deluxetable*}
\endgroup

\begingroup
\setlength{\medmuskip}{0mu}
\begin{deluxetable*}{cccccc}
  \tablecaption{Observed (\citetalias{Dai2023}) and Adjusted Period Ratios in TOI-1136}\label{tab:toi-1136}
  \tablecolumns{6}
  \tablehead{
    \colhead{} &
    \colhead{b:c} &
    \colhead{c:d} &
    \colhead{d:e} &
    \colhead{e:f} &
    \colhead{f:g}
  }
  \startdata
  \hline
    $P_2:P_1$        & $1.4995$   & $2.0008$   & $1.5016$   & $1.3999$   & $1.5024$   \\
    Commensurability & $3$:$2$    & $2$:$1$    & $3$:$2$    & $7$:$5$    & $3$:$2$    \\
    $\Delta$         & $-0.00031$ & $+0.00039$ & $+0.00107$ & $-0.00010$ & $+0.00163$ \\
  \hline
    \multicolumn{6}{c}{$\Delta_+$ values adjusted by hand to be $>0$ and used to initialize \texttt{Runs B-D}} \\
  \hline
    $\Delta_+$       & $+0.00100$ & $+0.00039$ & $+0.00107$ & $+0.00010$ & $+0.00163$ \\
  \hline
  \enddata
\end{deluxetable*}
\endgroup

\section{Long-Term Integrations Starting from Migration Simulations}
\label{sec:int_mig_sim}

Our first experiment is based on the disk migration simulations of \citet[][hereafter \citetalias{Dai2023}]{Dai2023}, intended to reproduce the orbital architecture of TOI-1136 (see our introduction).  These simulations (which adopt the first migration prescription in section 6.3 of \citetalias{Dai2023}; see also their Figure 10) modeled short-scale, Type-I (inward) migration and eccentricity damping of six planets, and yielded an ensemble of $N=160$ coplanar resonant chains following the parking of the innermost planet b at the inner disk edge.  The top panel of Figure \ref{fig:mig_sims} shows the distribution of period ratios $P_2/P_1$ in the 160 systems constructed across disk parameter space (varying disk mass and scale height and thus migration and eccentricity damping rates).  Most pairs are captured near first-order commensurabilities: 2:1, 3:2, 4:3, 5:4, and so on.  Note how the peaks near the 2:1 and 3:2 resonances lie preferentially at larger $P_2/P_1$ (i.e. $\Delta > 0$), as expected from eccentricity damping \citep[e.g.,][]{Choksi2020}.  A second-order 7:5 resonance is exhibited in 75 out of 160 chains, with 63  having the same sequence of resonances as in TOI-1136 (3:2, 2:1, 3:2, 7:5, 3:2).  Second-order resonances are generated by relatively slow migration in low-mass disks.

\begin{figure*}
  \centering
  \includegraphics[width=0.925\linewidth]{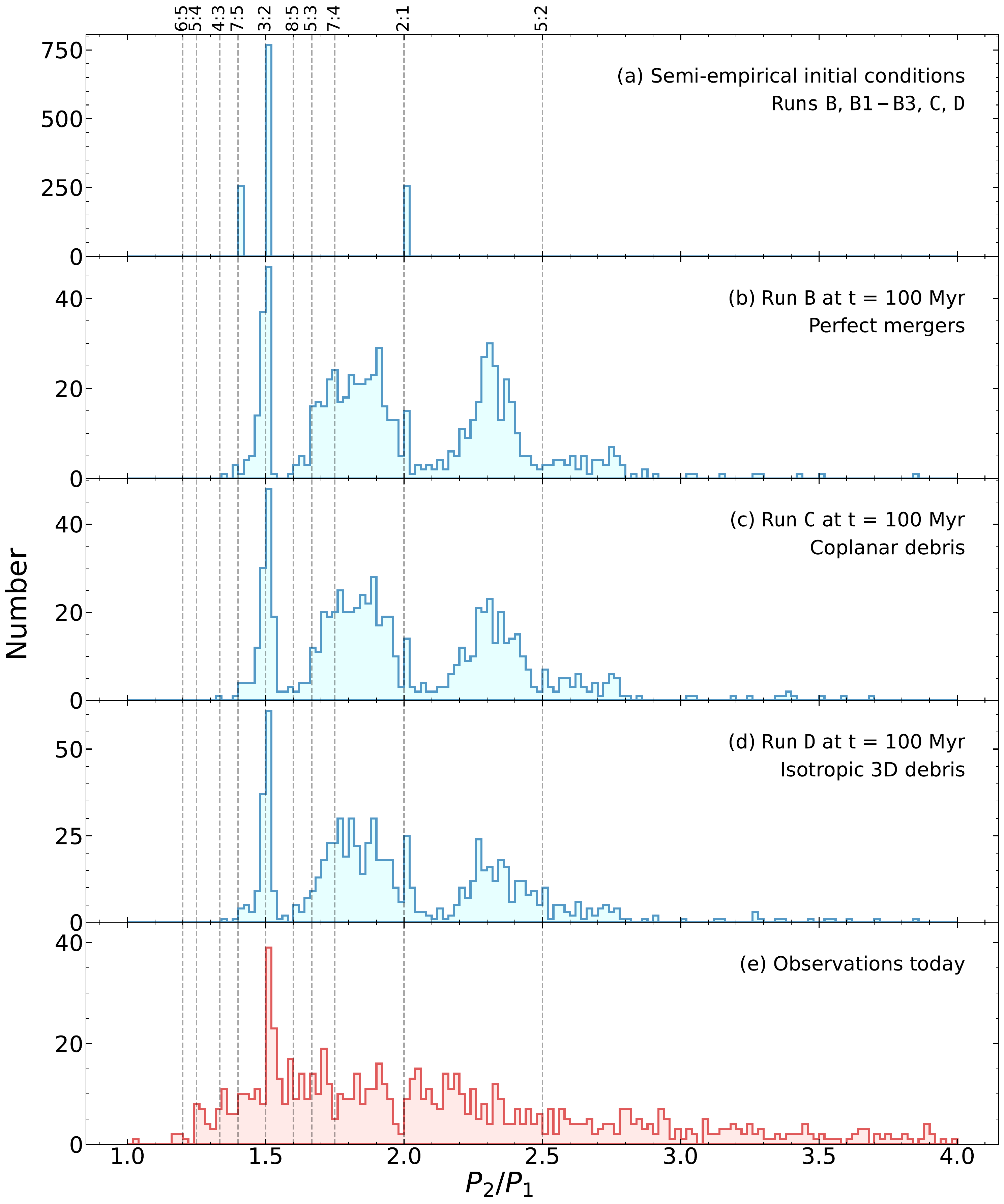}
  \caption{Neighboring planet period ratios
  for (a) our semi-empirical \texttt{Runs B-D} at the start of our integrations (note how all pairs are initialized with $\Delta_+ > 0$; see Table \ref{tab:toi-1136}); (b) the final state of \texttt{Run B}, which assumes collisions result in perfect mergers; (c) the final state of \texttt{Run C}, where major mergers produce coplanar debris particles that undergo scattering and minor mergers; (d) the final state of \texttt{Run D}, where major mergers produce debris particles ejected isotropically in 3D; and (e) the observations as downloaded for planets with radii $< 4R_\oplus$ from the NASA Exoplanet Archive in March 2024.  Positions of 1st, 2nd, and 3rd order resonances are shown with dotted vertical lines.  Note the absence of resonances between $P_2/P_1 = 1.8$ and 2.0; we propose that the observed ``trough'' just narrow of the 2:1 (and absent from the 3:2) arises from this desert.
  \label{fig:experiments5}}
\end{figure*}

We use the results of these 160 disk-migration simulations as initial conditions for our post-disk integrations (\texttt{Run A} in Table \ref{tab:runs}).  We integrate each six-planet system, disk-free, for 100 Myr using the \texttt{REBOUND} code \citep{rebound} outfitted with the hybrid-symplectic \texttt{MERCURIUS} integrator \citep{reboundmercurius} (similar results were obtained with the \texttt{TRACE} integrator; \citealt{Lu2024_trace}).  The integration timestep is set to $0.2$ days, approximately $5\%$ of the period of the innermost planet.  Note that \texttt{MERCURIUS} switches to an adaptive timestepper when there is a close encounter.  For planets like those in TOI-1136 whose orbital velocities exceed their surface escape velocities, dynamical instability and orbit crossing should lead to collisions and mergers, rather than ejections from the host star \citep[e.g.,][]{Goldreich2003}.  Unlike the original \citetalias{Dai2023} simulations, we allow for planetary collisions by assigning non-zero physical radii $R_{\rm p}$ to the planets (see Table 10 of \citetalias{Dai2023}).  In \texttt{Run A} (and also in the later \texttt{Run B} series), planetary collisions are assumed to result in perfect mergers conserving mass and momentum, with merger products assigned a new physical radius $R_{\rm merger}=(R_1^2+R_2^3)^{1/3}$.  Over the last $4.1$ years of the integrations (the \textit{Kepler} mission duration), we measure the orbital periods of all planets using mock transit observations.  We have verified in \texttt{Run A} and all other runs in our paper that angular momentum is conserved to better than 1\% (energy is not conserved in mergers).

As displayed in Figure \ref{fig:mig_sims}, the final distribution of neighboring-planet period ratios $P_2/P_1$ at $t = 100$ Myr is mostly unaltered from its initial distribution at $t=0$.  In particular the final period ratio distribution lacks the continuum of values observed for mature \textit{Kepler} systems (at $t \sim $ several Gyrs).  Mostly what happens over 100 Myr is the destruction of the closest resonant pairs (5:4 and narrower, and about half of the 4:3 population), which are unstable without eccentricity damping from the parent disk.  The planets initially captured into 3:2, 2:1, and 7:5 resonance largely stay there.

Our inability to reproduce the observed secular decline in resonance occupation (Fig.~\ref{fig:resonance_fraction}) using the migration simulations of \citetalias{Dai2023} implies that these simulations, which are typical of those in the literature, do not statistically represent the orbital configurations of young systems after disk dispersal.  For the remainder of this paper we try a different approach.

\begin{figure*}
  \centering
  \includegraphics[width=\linewidth]{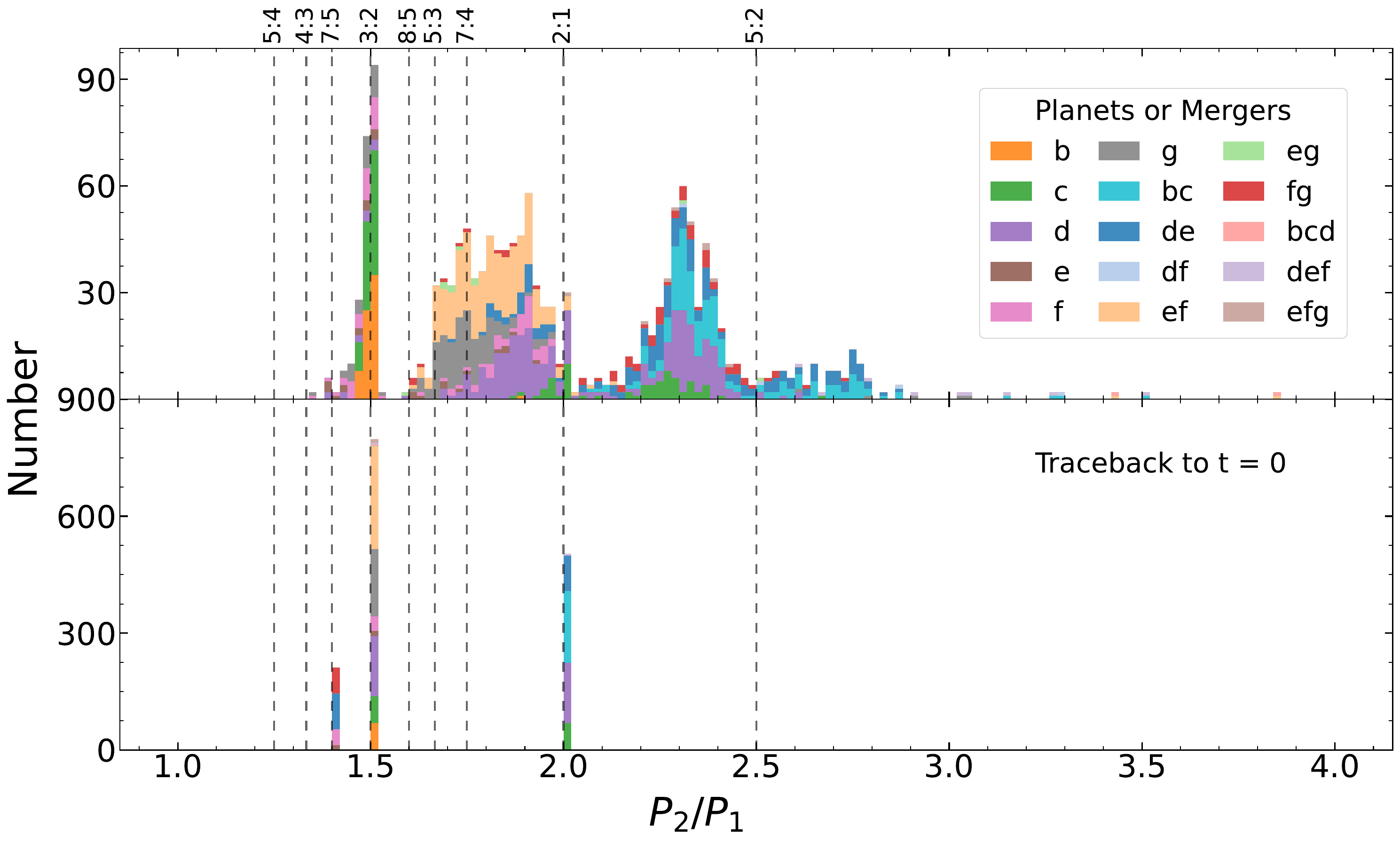}
  \caption{A history of violence: how planetary mergers alter period ratios.  Top panel: the final $P_2/P_1$ distribution at $t=100$ Myr for \texttt{Run B}, color-coded according to the planets that contribute to a given period ratio.  Combined letters denote a merger product; e.g., `bc' indicates the merged product of planets b and c, and 'bcd' the merger of planets b, c, and d.  Since every period ratio involves two planets, the histogram in the top panel is twice the height of the histogram for \texttt{Run B} as shown in the second panel of Figure \ref{fig:experiments5}.  Bottom panel: tracing the planet pairs contributing to a given period ratio at $t = 100$ Myr back to their original period ratios at $t=0$.  If a pair at $t = 100$ Myr involves merger products, we assign both members of the pair to the original period ratio that connects the pair; thus, e.g., a pair consisting of planet d and merger ef in the top panel are mapped to the original period ratio between d and e ($3/2 + \Delta_+$) in the bottom panel, and not the ratio between e and f ($7/5 + \Delta_+$).  In this way, we have a one-to-one correspondence between the two panels, excepting the rare mergers of non-neighboring planets (4 df and 10 eg planets) for which assigning unique initial period ratios is too ambiguous.
  \label{fig:decomposed_P2overP1}}
\end{figure*}

\begin{figure}
  \centering
  \includegraphics[width=\linewidth]{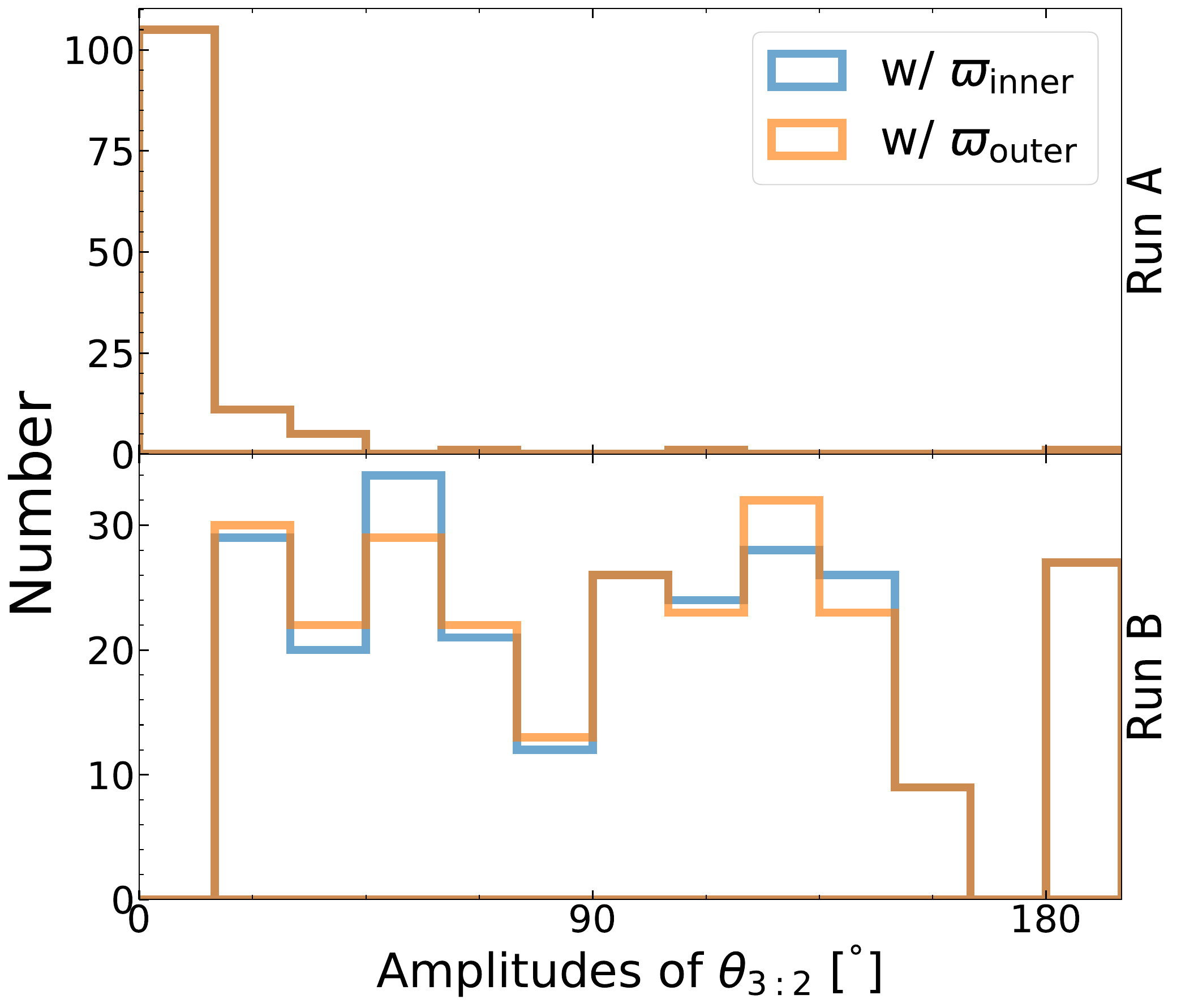}
  \caption{Initial 3:2 resonant libration amplitudes of the innermost b:c planet pairs for \texttt{Run A} (top panel) and \texttt{Run B} (bottom panel).  There are two resonant arguments for the co-planar 3:2 resonance, one which depends on the longitude of periapse of the inner planet b (blue histogram), and another which depends on that for the outer planet c (orange histogram).  Libration amplitudes are determined by sinusoidal fits to the resonant argument vs.~time, and evaluated as the deviation from libration center.  In the top panel, only the 124 systems (out of a total of 160 in \texttt{Run A}) that have the b:c pair near a 3:2 resonance are plotted, and of these only 1 system is in circulation, which we assign a libration amplitude of 180$^\circ$.  For the observation-based simulations in \texttt{Run B}, 27 out of 256 systems are in circulation, including those that switch between libration and circulation.  The larger initial libration amplitudes in \texttt{Run B} explain why its resonant chains break more easily than those in \texttt{Run A}.
  \label{fig:theta_32}}
\end{figure}

\section{Long-Term Integrations Starting from TOI-1136}
\label{sec:int_empirical}

The null results from the previous section imply that there must be physical effects missing from standard disk-migration and eccentricity damping simulations in assembling and shaping resonant chains.  There does not seem to be consensus as to what these are; candidates include stochastic torques from disk turbulence (e.g.~\citealt{Rein2012}; \citealt{Goldberg2023}), mass loss from planetary gas envelopes (\citealt{Matsumoto2020}; \citealt{Owen2024}), overstable librations from eccentricity damping (\citealt{Goldreich_Schlichting_2014}; \citealt{Nesvorny2022}), or scattering and collisions from planetesimals (\citealt{Chatterjee2015,Ghosh2023}; \citealt{Wu2024}).  Rather than investigate this issue from first principles, we take a different tack, leveraging the observations in an admittedly ad hoc and arbitrary manner to explore possible post-disk initial conditions.  We rely again on TOI-1136 as a guide, but now construct an ensemble of six-planet systems by uniformly drawing orbital elements within the one-sigma uncertainties of TOI-1136's observed best-fit parameters (\citetalias{Dai2023}, Table 10)  These elements cover substantially more parameter space than the simulation-based elements in \texttt{Run A}, and are assumed to embody whatever (unknown) perturbations stirred up resonant chains to the point of breaking.  The goal is not to decide what physical effects are responsible for breaking chains, but to study how they break if they break, and how observations may be reproduced.  The philosophy is similar to that of \citet{Goldberg_Batygin_2022}; we will recover similarly positive results, and fill in more details regarding resonant sub-structures in the distribution of period ratios.

We modify this observation-based set of initial conditions in one respect: where $\Delta < 0$, we change the orbital period ratio to $\Delta > 0$.  We do this for two reasons.  First, disk migration with eccentricity damping predicts planets resonantly capture to $\Delta > 0$, for first-order resonances in a time or ensemble-averaged sense \citep[e.g.,][and our Figure \ref{fig:mig_sims}]{Choksi2020}.  Second, we found upon experimentation that when our initial conditions follow the preference for $\Delta < 0$ as per the \citetalias{Dai2023} observations of TOI-1136, the final $\Delta$ distribution after integration preserves the preference for $\Delta < 0$, contrary to observation.

Thus, for the innermost 3:2 b:c pair, we do not use the observed best-fit $\Delta = -0.00031$, but instead use $\Delta_+ = +0.001$ (similar to the observed $\Delta = +0.00107$ for the other 3:2 d:e pair in the system; see Table \ref{tab:toi-1136}).  We make a similar replacement for the 7:5 e:f pair, flipping the sign of the initial $\Delta$ from -0.0001 to +0.0001 (though capture into second-order resonance has not been shown to favor $\Delta > 0$).  To calculate the initial orbital periods (semi-major axes) for the six planets, we first set the orbital period of planet b to its observed best-fit value of 4.17278 days (\citetalias{Dai2023} Table 10), and then use the $\Delta_+$ values in our Table \ref{tab:toi-1136} to determine the remaining orbital periods.  All other orbital parameters (planet-to-star mass ratio, mean anomaly, eccentricity, longitude of pericenter) are sampled independently and within the ranges listed in Table 10 of \citetalias{Dai2023}.  Our independent sampling ignores correlations between fitted elements, and thus surveys a large volume of parameter space in which there is no guarantee of stability.

Altogether we construct $N = 256$ coplanar TOI-1136 clones for 100-Myr integrations with \texttt{MERCURIUS}, assuming perfect mergers --- this experiment constitutes our \texttt{Run B}.  For this run and all subsequent runs in this paper, we adopt a default timestep of 0.145 days $\simeq3\%$ the period of the innermost planet.  Shorter timesteps are taken by \texttt{MERCURIUS} during close encounters.

The top two panels of Figure \ref{fig:experiments5} show the initial and final $P_2/P_1$ distributions.  In contrast to \texttt{Run A}, many of the resonant chains in \texttt{Run B} break.  They break when a pair of planets stir each other to sufficiently large eccentricity that they collide and merge.  The merger product tends to have an orbital period intermediate between that of the colliders, widening adjacent orbital spacings.  Thus, if planet e collides with f, in addition to the e:f = 7:5 resonance being destroyed, the resultant period ratio between d and the merger product ef increases from the original d:e=3:2,  and similarly the period ratio between the merger product ef and g increases from the original f:g = 3:2.  In this way pairs initially in the 3:2 resonant peak are displaced to a broader continuum of values $P_2/P_1 > 3/2$ in Figure \ref{fig:experiments5} --- see the light orange blocks marked `ef' in Figure \ref{fig:decomposed_P2overP1}, which pair with either the dark purple blocks marked `d' or the grey blocks marked `g'.  Similarly, a second continuum of $P_2/P_1 > 2$ ratios is created from mergers of b and c.

Intriguingly, and perhaps by chance, the first continuum falls just short of $P_2/P_1 = 2$, creating a ``peak-trough'' structure near the 2:1 resonance like the one observed for mature systems.  The initial 7:5 pairs appear largely eliminated, also in agreement with {\it Kepler} statistics.  One problem is that some of the b:c = 3:2 pairs that survive appear to have evolved to $\Delta < 0$, in a proportion that seems too large to compared to observations of mature systems.

Dynamical instabilities in \texttt{Run B}, and their absence in \texttt{Run A}, can be attributed to differences in their initial resonant libration amplitudes, as shown in Figure \ref{fig:theta_32} for the innermost b:c = 3:2 pairs.  Larger libration amplitudes in \texttt{Run B} stem from the observational uncertainties on which this run is based (see also \citealt{Jensen2022}), and permit closer and more varied encounters between planets.  For a deeper analysis of the mechanics of instability in chains, see \citet{Goldberg_etal_2022}.

\subsection{Further Experiments Assuming Perfect Mergers (\texttt{Runs B1-B3})} \label{sec:b1_b3}

\begin{figure*}
  \centering
  \includegraphics[width=0.96\linewidth]{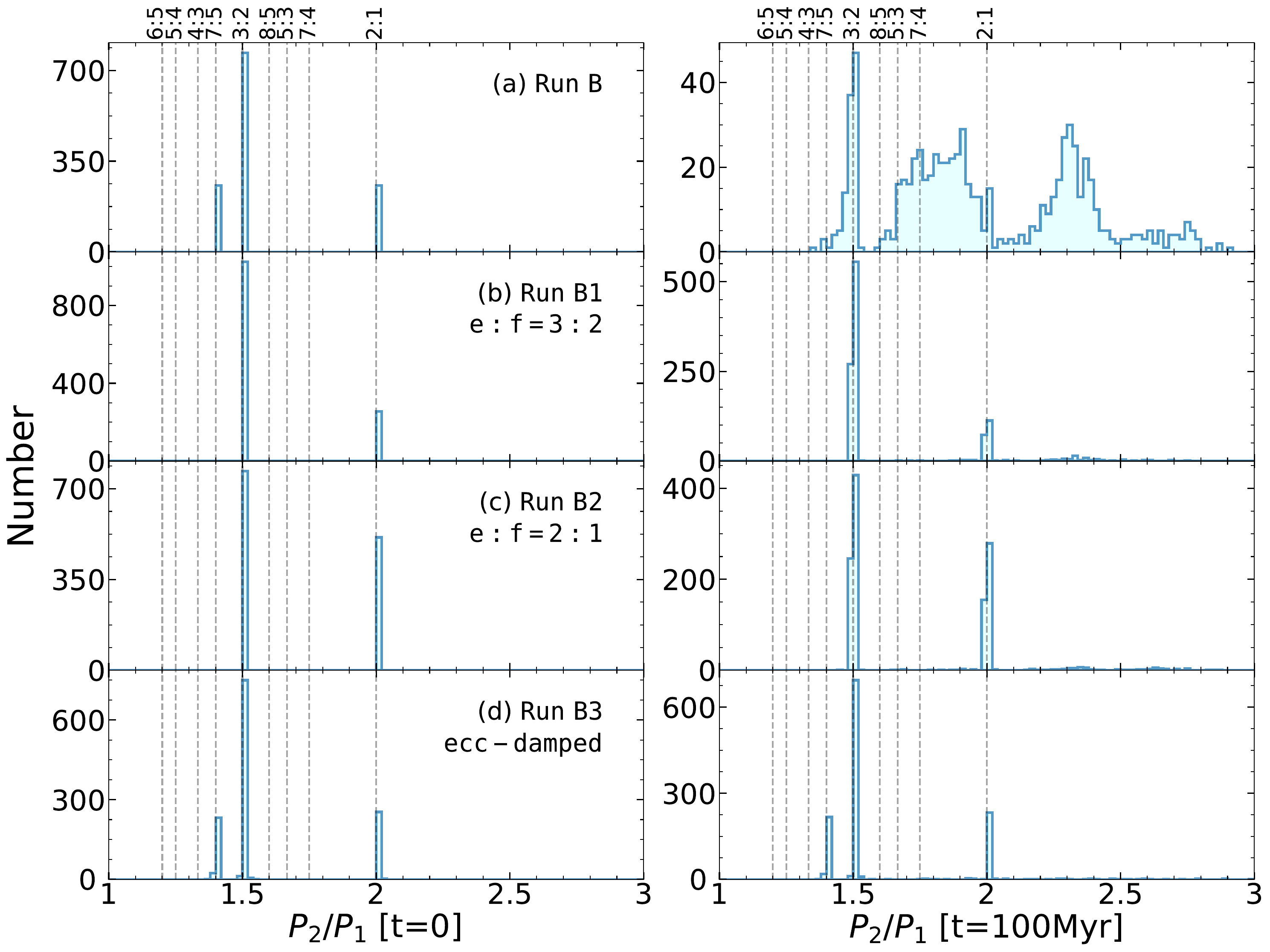}
  \caption{Initial (left column) and final (right column) period ratio distributions for \texttt{Runs B, B1, B2,} and \texttt{B3} (top to bottom rows; see also Table \ref{tab:runs}).  These experiments demonstrate that the presence of a weak 2nd-order resonance like the 7:5 (absent in \texttt{Runs B1-B2}), and sufficiently large initial eccentricities (absent in \texttt{Run B3}), help promote instability and the breaking of resonant chains.
  \label{fig:runB_series}}
\end{figure*}

\texttt{Runs B1-B3} are variations on \texttt{Run B}.  In \texttt{Run B1} we replace the original e:f = 7:5 resonance with the 3:2 resonance, changing only the period ratio for e:f (and not changing $\Delta_+$, which remains at 0.0001).  All other parameters (eccentricity, mean anomaly, etc.) are sampled the same way as in \texttt{Run B}. \texttt{Run B2} is a similar experiment that replaces the e:f 7:5 resonance with the 2:1 resonance.  As shown in Figure \ref{fig:runB_series} (second and third panels from the top), the replacement of the 2nd order 7:5 resonance with a 1st order resonance appears to stabilize the system: apart from some systems spreading to $\Delta < 0$, there are hardly any mergers.  It may be that 2nd order resonances, which are intrinsically weaker than 1st order resonances, play a prominent role in breaking resonant chains over time (see also section 7.2 of \citetalias{Dai2023}, and Xu \& Lai 2017).

In \texttt{Run B3}, before we integrate for 100 Myr, we damp eccentricities using the \texttt{modify\_orbits\_forces} routine in \texttt{REBOUNDx} \citep{Tamayo2020} and integrate for $10\tau_e = 10 e/\dot{e}$, so that all planets have new initial eccentricities $< 0.01$ (by comparison, in \texttt{Run B}, initial eccentricities range from $\sim$0.01 to $\sim$0.12; \citetalias{Dai2023} Table 10).  The resultant systems are completely stable over 100 Myr (Figure \ref{fig:runB_series} bottom panel).  Sufficiently large eccentricities---technically free eccentricities that give rise to non-zero resonant libration amplitudes---appear essential for breaking resonant chains.

\subsection{Imperfect, Debris-Generating Mergers (\texttt{Runs C-D})}
\label{sec:minor}

Mergers are messy, as recently emphasized by \citet[][and references therein]{Wu2024}.  Collisions at relative velocities greater than the planets' surface escape velocities are mostly oblique and hit-and-run \citep[e.g.,][]{Aarseth1993, Leinhardt2012, Cambioni2019, Emsenhuber2019,  Emsenhuber2020}, producing debris not immediately re-accreted by the colliders. \citet{Esteves2022} simulated imperfect mergers in breaking chains, finding the gross evolution to be largely the same as when mergers are assumed perfect.  We do not disagree, but here focus on the effects that debris and imperfect mergers have on the finer resonant sub-structures in the period ratio distribution.  \citet{Wu2024} have shown that scattering of planetesimals strewn between and around a pair of resonant planets can wedge them farther apart, potentially explaining the preference for $\Delta > 0$ observed among mature sub-Neptunes.

Accordingly in \texttt{Runs C} and \texttt{D}, we relax the assumption of perfect mergers.  Each time two planets collide, the merger product is assumed to contain 90\% of the combined mass, with the remaining 10\% distributed equally among $N_{\rm debris} = 30$ new particles.  Debris particles carry mass but do not interact with each other; they only interact with planets via gravity, and can collide with planets in subsequent ``minor mergers'' that are assumed perfect.  The debris particles are born with isotropic ejection velocities.  Their initial locations are randomly displaced by $\bm{\delta r}$ from the collision site, where $|\delta r|$ is 50$\times$ the physical radius of the merger product, and their initial velocities are those of the merger product plus a velocity $\bm{\delta v}$ parallel to $\bm{\delta r}$, with $|\delta v|$ equal to $30\%$ of the relative impact velocity between merger progenitors.  The above prescription is made more for computational ease than anything else, though the fraction of mass ejected (10\%) and the ratio of ejection to collision velocities (30\%) do not seem unrealistic for the mass that escapes the gravity well of the merger product (e.g.~\citealt{Esteves2022}; see also Figure 3 of \citealt{Gabriel_Cambioni_2023}, and references therein).  To ensure the collision does not alter the planetary system's center of mass or angular momentum, debris particles are created in diametrically opposed pairs (with $\pm \bm{\delta r}$).  In \texttt{Run C}, debris particles are restricted to be coplanar with the planetary system ($\bm{\delta r}$ traces a circle).  In \texttt{Run D}, we allow for 3D motions ($\bm{\delta r}$ traces a sphere).  For each of \texttt{Runs C} and \texttt{D}, initial conditions for $N=256$ TOI-1136 clones are identical to those in \texttt{Run B} (i.e. the same seed random numbers were used to sample initial conditions for all \texttt{Runs B-D}).

\begin{figure*}
  \includegraphics[width=0.495\linewidth]{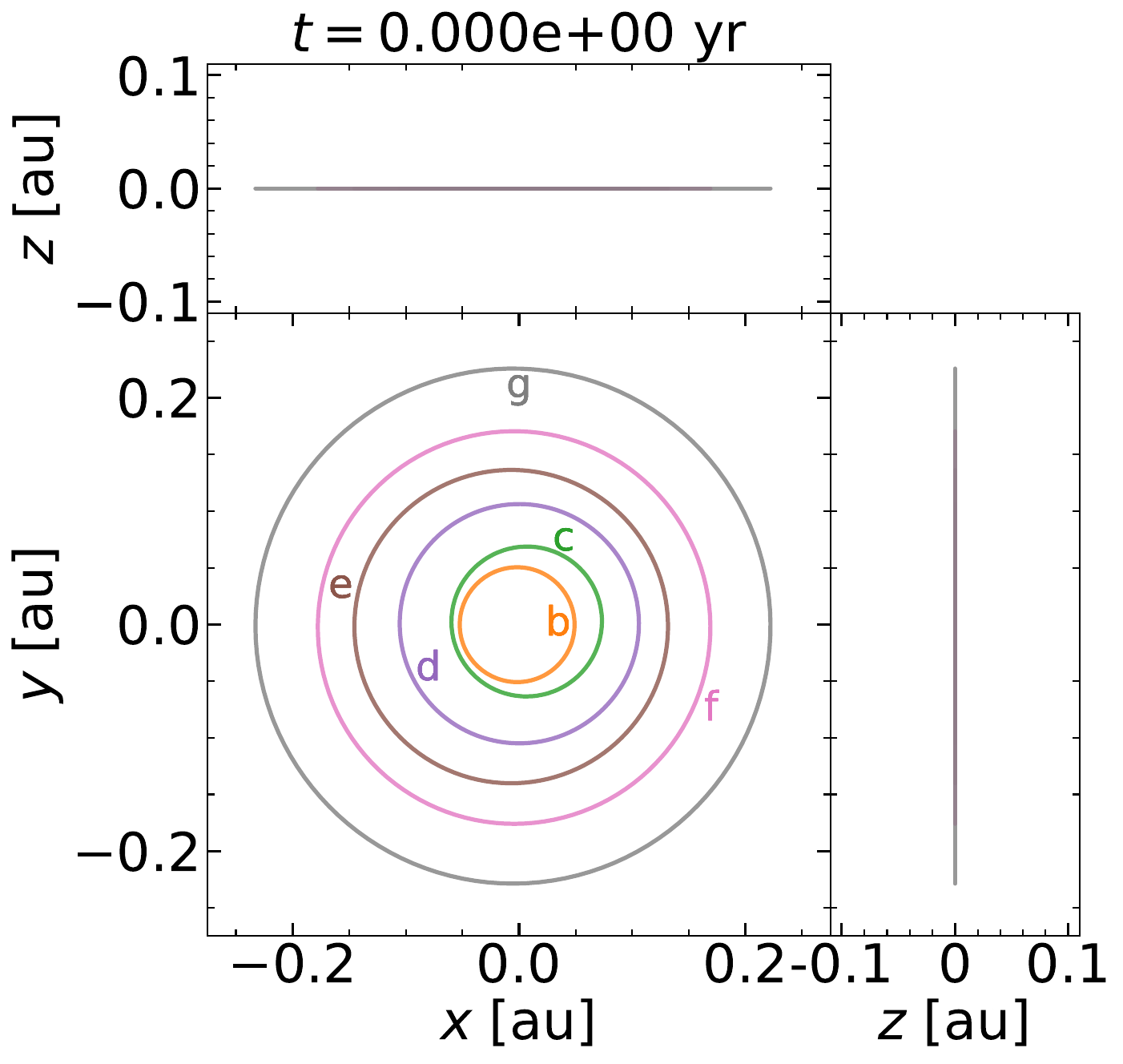}
  \includegraphics[width=0.495\linewidth]{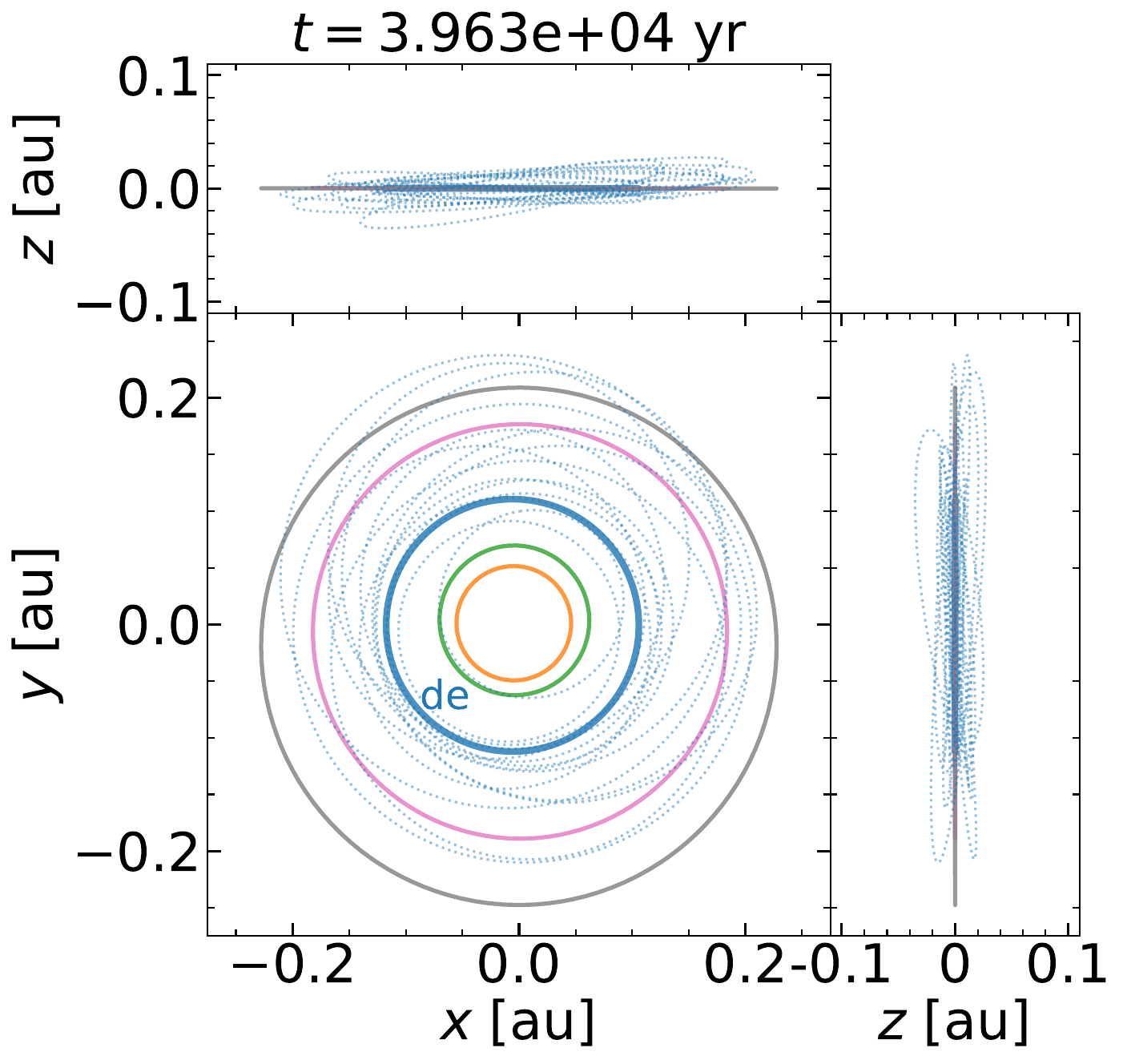}
  \vskip 1em
  \includegraphics[width=0.495\linewidth]{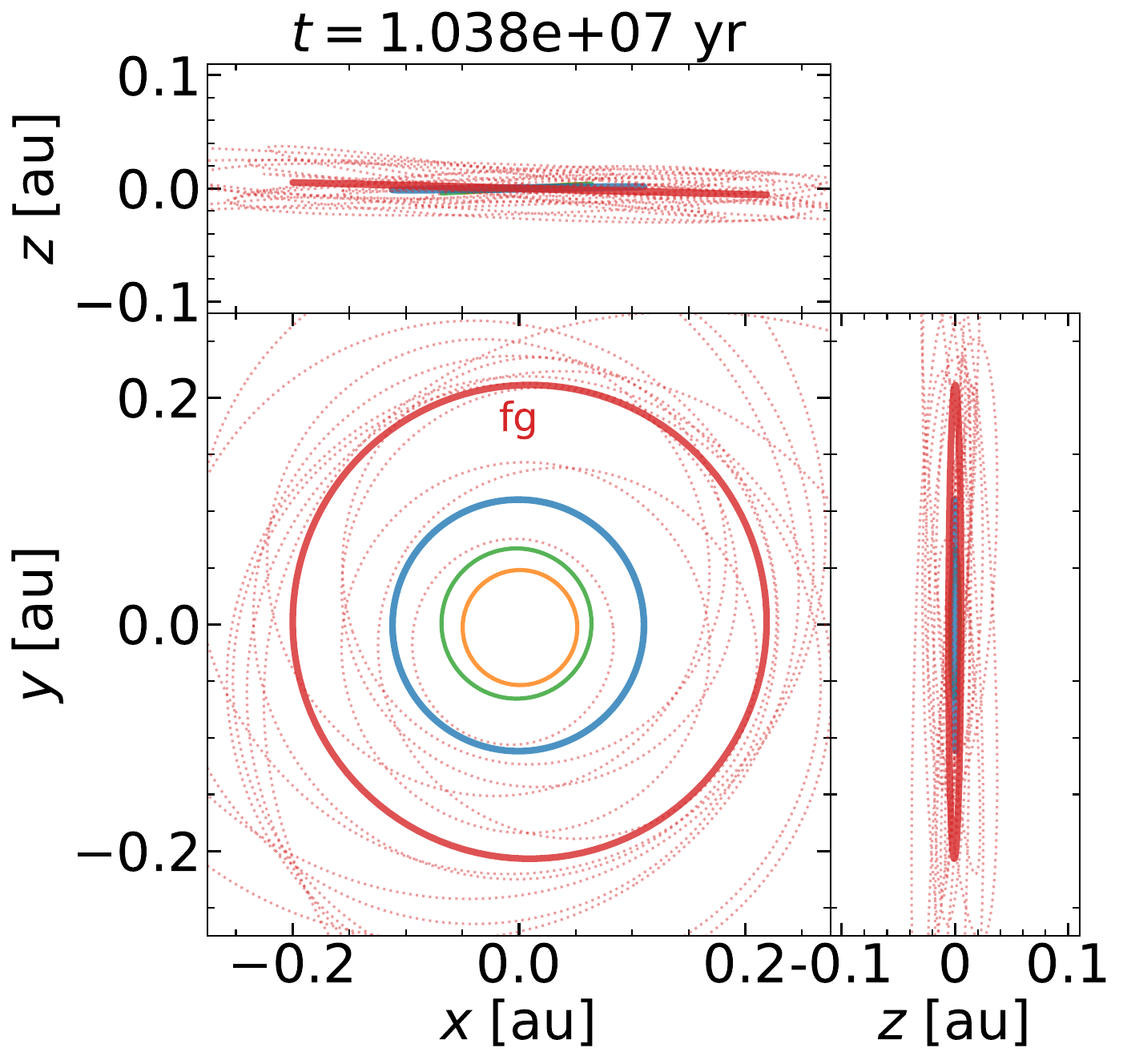}
  \includegraphics[width=0.495\linewidth]{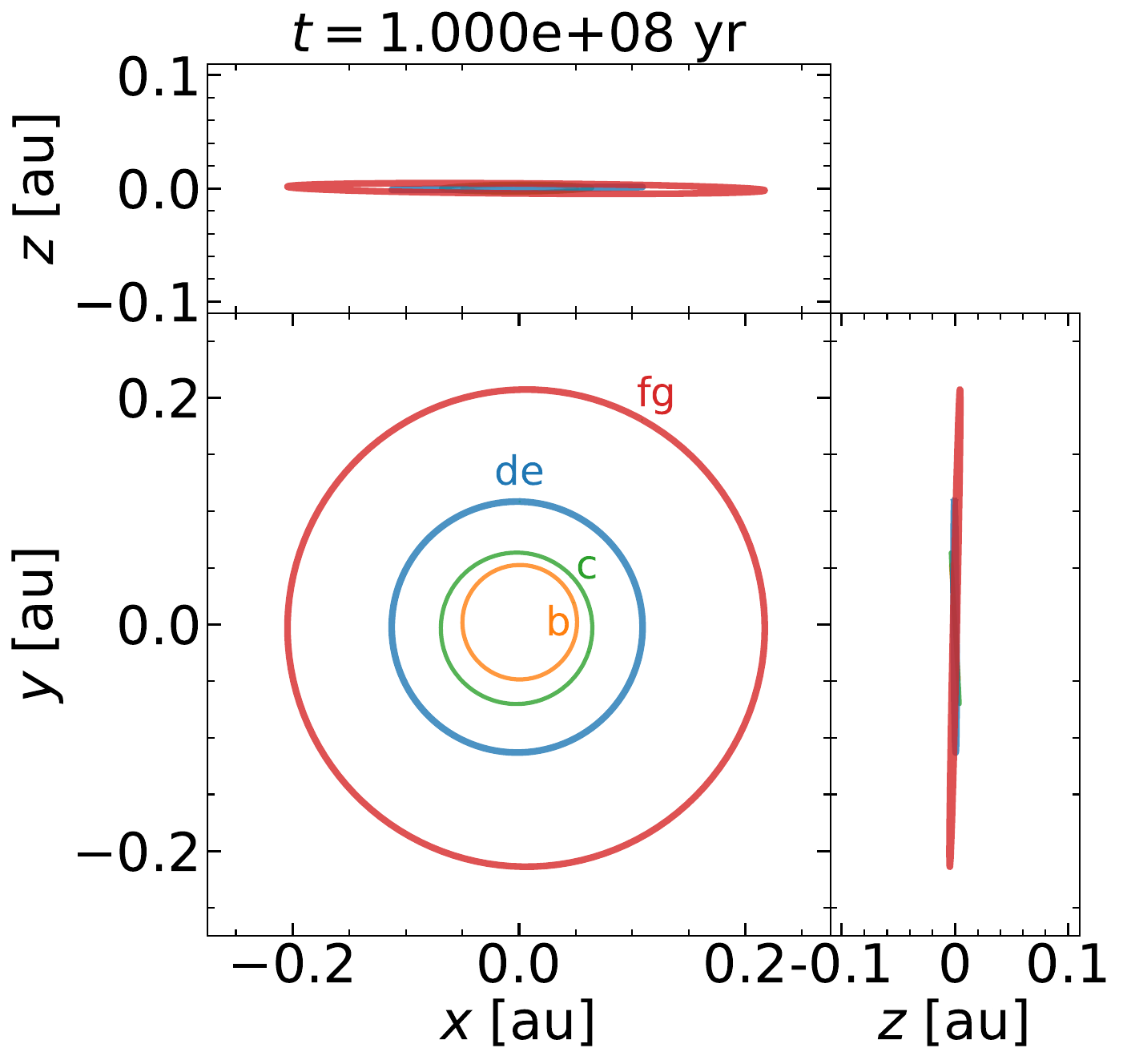}
  \caption{Snapshots seen in various $x$-$y$-$z$ projections of one example simulation from \texttt{Run D}, featuring two major merger events at $3.888\times10^4$ yr (creating planet de) and $1.038\times10^7$ yr (creating planet fg).  The orbits of the initial planets (b, c, d, e, f, and g) are shown in thin solid lines, the orbits of merger products (de, fg) are drawn in thick solid lines, and the orbits of debris particles are plotted in dotted lines with colors that match those of their parent merger products.
  \label{fig:snapshots1_RunD}}
\end{figure*}

\begin{figure}
    \centering
    \includegraphics[width=\linewidth]{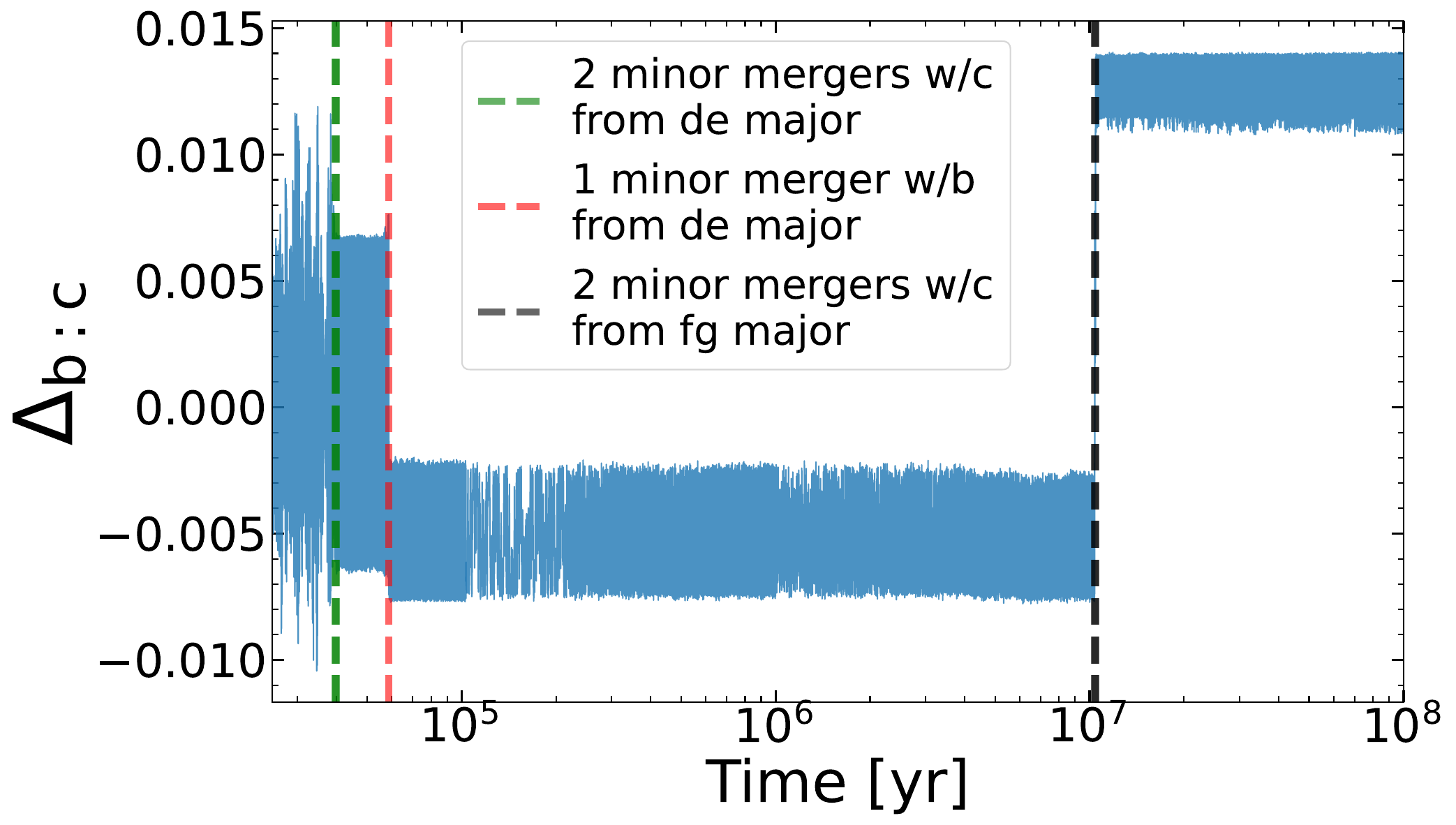}
    \caption{The evolution of $\Delta$ for the innermost b:c = 3:2 planet pair in the \texttt{Run D} simulation shown in Figure \ref{fig:snapshots1_RunD}.  Minor mergers alter $\Delta$, lowering it when a debris particle collides with b (green dashed line), and later raising it when another debris particle collides with c (black dashed line).  For the final transit timing variations (TTVs) of this pair, see Figure \ref{fig:good_TTV_phase}.
    \label{fig:Delta_bc}}
\end{figure}

\begin{figure}
    \centering
    \includegraphics[width=\linewidth]{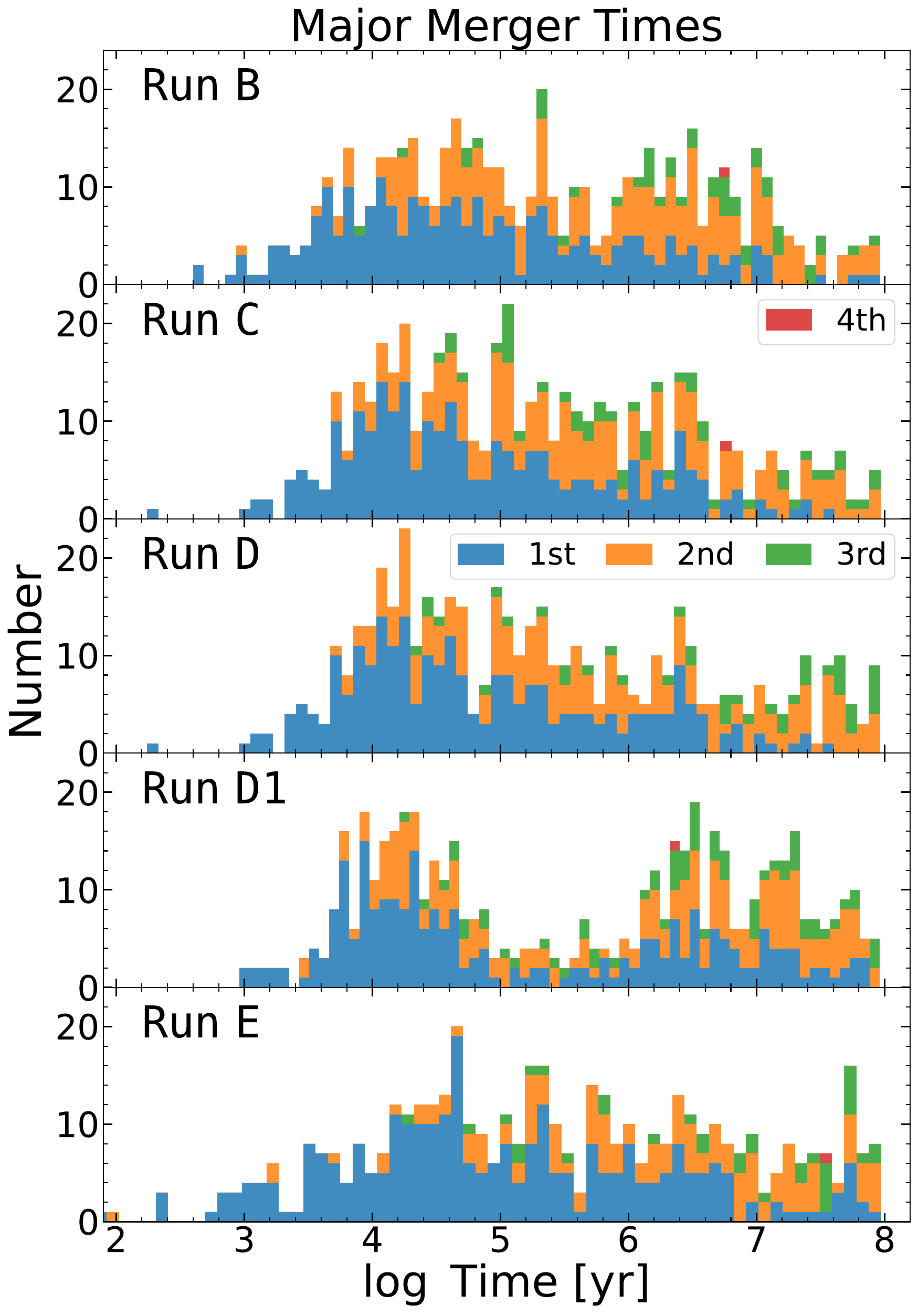}
    \caption{Times when major mergers occur in \texttt{Runs B-E}, color-coded according to whether a merger occurred chronologically 1st, 2nd, 3rd, or (in rare cases) 4th within a given system.  Too many resonance-breaking mergers occur at times $< 10^6$ yr to be consistent with the observed decline in resonance fraction over $10^8$ yr timescales (Figure \ref{fig:resonance_fraction}).
    \label{fig:merger_time}}
\end{figure}

\begin{figure}
    \centering
    \includegraphics[width=\linewidth]{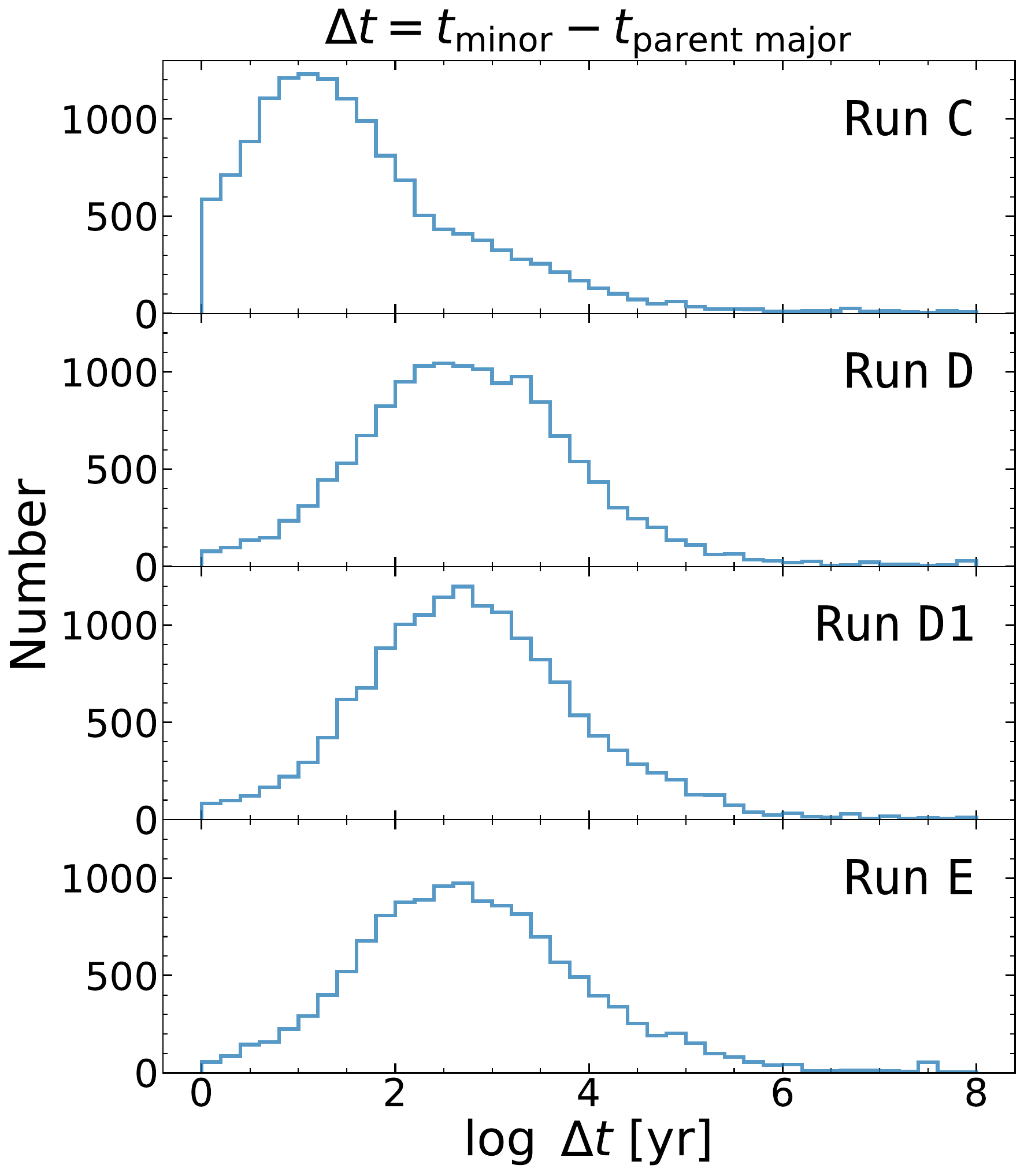}
    \caption{Minor mergers between debris particles and planets occur mostly $\Delta t < 10^4$ yr after the debris particles are born in parent major merger events.  Debris particles in \texttt{Runs D, D1,} and \texttt{E} are not coplanar with planets and take longer to accrete than in \texttt{Run C} where they are coplanar.
    \label{fig:t_minor_from_t_Major}}
\end{figure}

\begin{figure}
  \centering
  \includegraphics[width=\linewidth]{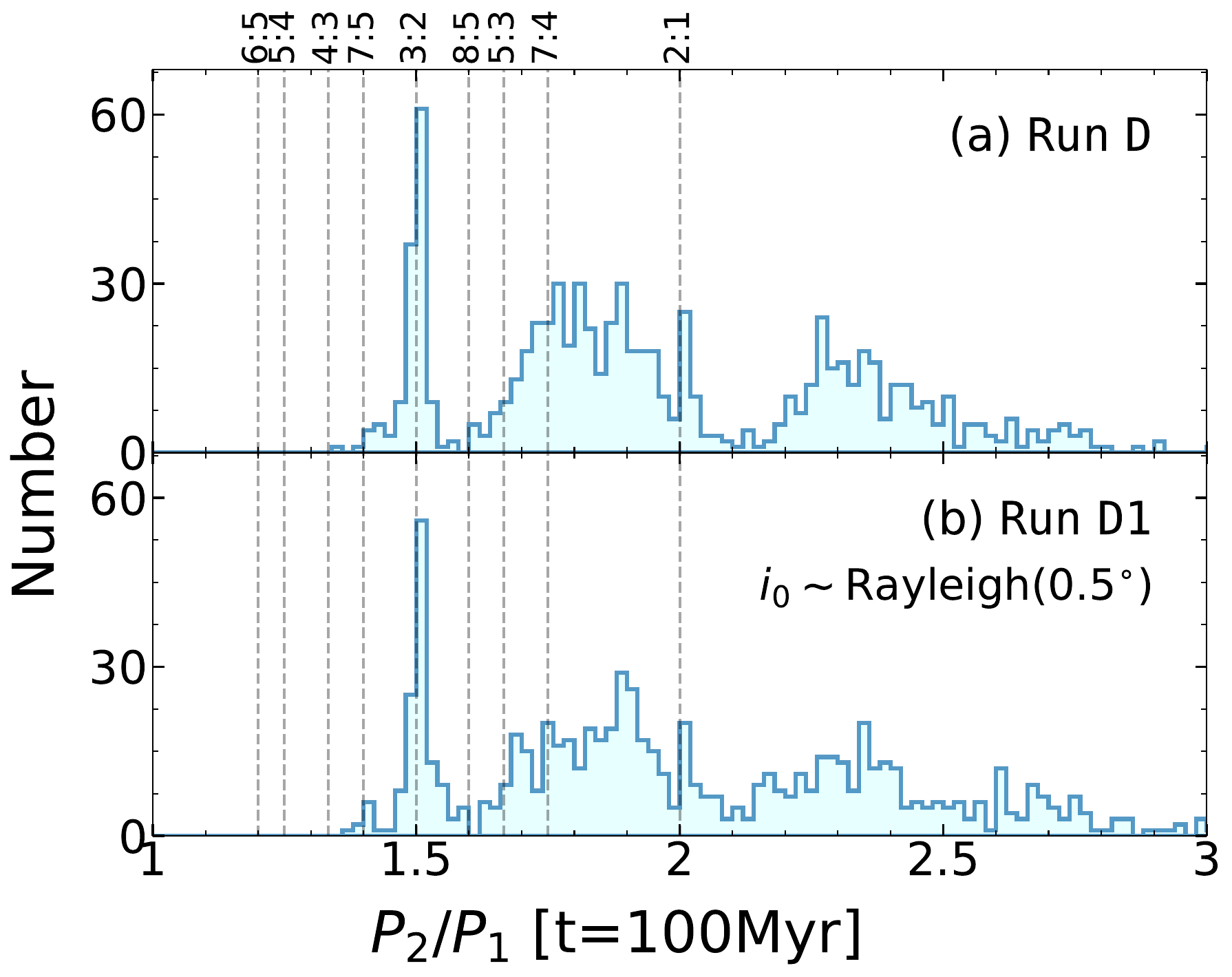}
  \caption{Comparison of final period ratio histograms between \texttt{Run D} with initially coplanar planets and \texttt{Run D1} with initially non-coplanar planets (see Table \ref{tab:runs}).  Both runs exhibit the 2:1 trough, while the non-resonant continuum is smoother and less bimodal in \texttt{Run D1}, better matching observations.
  \label{fig:runD_series}}
\end{figure}

Figure \ref{fig:experiments5} compares the post-integration $P_2/P_1$ ratios for \texttt{Runs B-D}.  The results are similar, with \texttt{Run D} arguably yielding a peak-trough 2:1 asymmetry more closely resembling that observed today.  The continuum of values at $3/2 < P_2/P_1 < 4$ also slopes more gently downward with increasing period ratio, in a manner similar to the observations, modulo the bimodality of the simulations (see discussion of \texttt{Run B} preceding section \ref{sec:b1_b3}).  The simulated 3:2 peaks all seem to have somewhat too many systems at $\Delta < 0$ compared to observations, though this problem appears minimized for \texttt{Run D}.  Following the definition of first-order resonance adopted in \citet[][i.e., $-0.015 < \Delta < +0.03$]{Dai2024}, the fraction of first-order resonant pairs is $15.5\%$ for \texttt{Run B}, $16.5\%$ for \texttt{C}, and $20.8\%$ for \texttt{D} at $t = 100$ Myr.  These fractions are comparable to the observed fraction of $\sim$25\% for ages $100$ Myr -- $1$ Gyr, as shown in Figure \ref{fig:resonance_fraction}.  About 1\% of the simulated systems survive unscathed to reproduce the same sequence of resonances as in TOI-1136.

Figure \ref{fig:snapshots1_RunD} shows four  snapshots of one of the simulations in \texttt{Run D}.  Once created, the debris particles scatter about.  Most eventually merge with the parent merger product, but a fraction are cast more widely.  In the example shown, of the 30 debris particles created when planets d and e merge, 21 are accreted by de, 1 collides with b, 2 with c, 4 with f, and 2 with g.  Of the 30 debris particles created when f and g merge, 25 are accreted by fg, 3 collide with de, and 2 with c.  These minor mergers tend to widen orbital spacings between resonant pairs, insofar as the debris particles just graze either member of the pair from the outside (which they do not always do).  For example, a debris particle created from the merger of f and g has its eccentricity increased by scattering until its orbit just intersects the orbit of planet c; the subsequent merger with c widens c's orbit by of order 1\% (the mass ratio between the debris particle and the planet) and thus increases $\Delta$ for the b:c 3:2 resonance---see Figure \ref{fig:Delta_bc} which shows this evolution, but which also shows that minor mergers do not always unfold this way.  Such widening of resonant pairs from ``outside grazers'' differs from, but complements, the widening mechanism of \citet{Wu2024} from planetesimal scattering.

The timestamps in Figure \ref{fig:snapshots1_RunD} point to a problem with our simulations, which is that the resonances tend to break too soon compared to the $\sim$$10^8$ yr timescale over which actual systems are observed to break (Fig.~\ref{fig:resonance_fraction}).  This is a generic problem with \texttt{Runs B}, \texttt{C}, and \texttt{D}.  Figure \ref{fig:merger_time} shows that more than half of the major mergers for these runs, out to the fourth merger in a given system, occur at times $< 10^7$ yr.  Most systems experience just two mergers, with half of these occurring at $< 10^6$ yr.  At $t = 10^6$ yr, the fraction of systems exhibiting at least one 1st or 2nd-order resonance is 47.6\%, and the fraction of pairs in 1st-order resonance is 39.6\%.  These fractions are too low compared to those shown at the earliest times in Fig.~\ref{fig:resonance_fraction}---this despite our initial conditions being drawn from \citetalias{Dai2023} which ostensibly screened out orbits that were unstable on timescales $< 1$ Myr (their section 6.1).  Probably the appearance here of such short-lived orbits stems from our ignoring correlations when sampling the orbital elements from Table 10 of \citetalias{Dai2023}.  Minor mergers add negligibly to major merger times---see our Figure \ref{fig:t_minor_from_t_Major}.  To summarize, the dynamical instabilities in our simulations unfold too rapidly to be consistent with Figure \ref{fig:resonance_fraction}, which shows $\sim$70\% of observed pairs are still in 1st-order resonance at ages as old as $10^7$--$10^8$ yr.

\subsection{Non-coplanar Planets and Mergers (\texttt{Run D1})}
\label{sec:d1}

Planets at the end of \texttt{Run D} have non-zero mutual inclinations after interacting with collisional debris ejected isotropically in 3D.  Here we experiment further with non-coplanarity by initializing the planets with non-zero mutual inclinations.  Observations suggest mutual inclinations on the order of $\sim$1$^\circ$ \citepalias{Dai2023}.  In \texttt{Run D1}, we draw planet inclinations randomly from a Rayleigh distribution with dispersion $\sigma = 0.5^\circ$. (Experiments with larger $\sigma = 3^\circ$, with initial inclinations comparable to initial eccentricities, were too strongly destabilizing, resulting in excessive mergers and no surviving resonant pairs.)

Figures \ref{fig:merger_time}, \ref{fig:t_minor_from_t_Major}, and \ref{fig:runD_series} show the results from \texttt{Run D1}.  They appear largely the same as \texttt{Run D}, except for a smoother and less bimodal continuum of period ratios after 100 Myr, and more prolonged times for major mergers.  Both changes better match observations.

\section{Long-Term Integrations Starting From Diverse Chains}
\label{sec:random_experiment}

\begin{figure*}
  \centering
  \includegraphics[width=\linewidth]{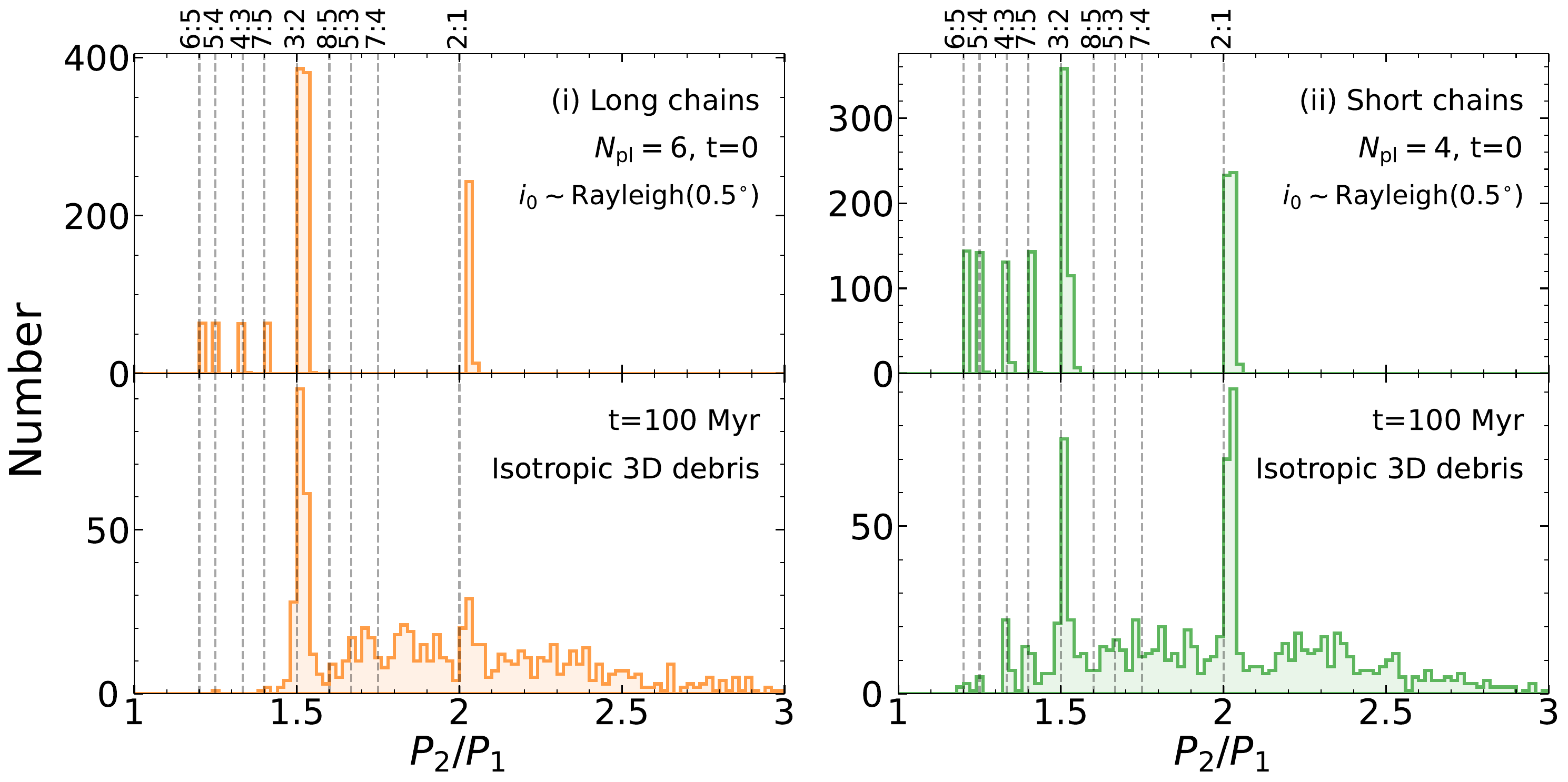}
  \caption{A library of diverse chains and their outcomes after 100 Myr of dynamical evolution.  Initial (top) and final (bottom) neighboring planet period ratios for type (i) chains initially containing six planets (left column) and type (ii) chains initially containing four planets (right column).
  \label{fig:long_short_chains}}
\end{figure*}

\begin{figure*}
  \centering
  \includegraphics[width=\linewidth]{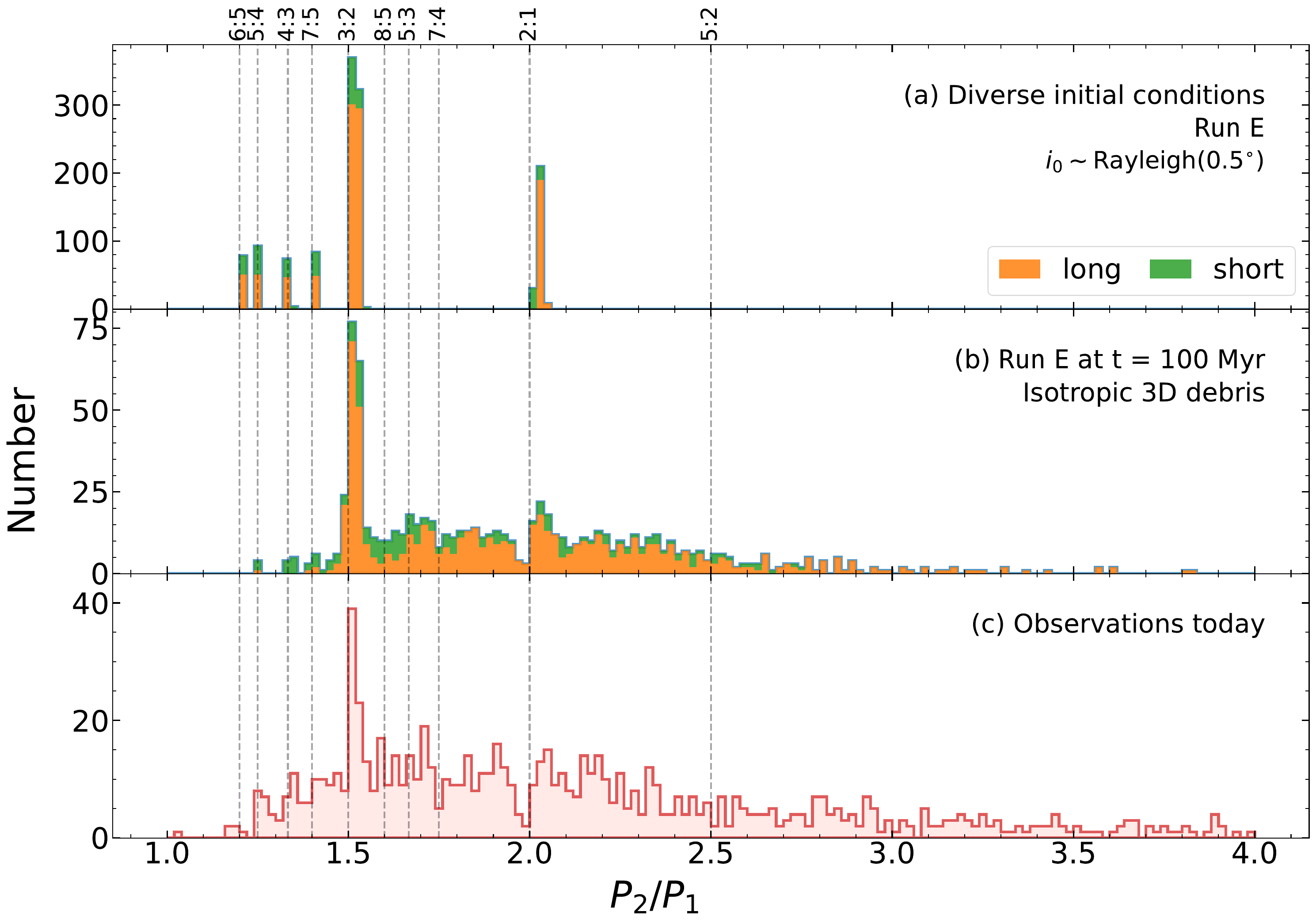}
  \caption{Neighboring planet period ratios for (a) \texttt{Run E} before integration, (b) \texttt{Run E} after 100 Myr of integration, and (c) observed Gyr-old planets with radii $< 4R_\oplus$ in the March 2024 NASA Exoplanet Archive.  In the upper two panels, histograms are color-coded according to whether a long chain (6-planet, type i) or short chain (4-planet, type ii) contributes.  Short chains (green) fill in the period ratio gap just wide of the 3:2 left open by long chains (orange).  The prominent 2:1 trough arises predominantly from the dissolution of long chains.
  \label{fig:random_exp}}
\end{figure*}

The experiments of Section \ref{sec:int_empirical} reproduce some features of the modern-day $P_2/P_1$ histogram, notably the peak-trough asymmetry near the 2:1 resonance.  However, the non-resonant portions of the simulated period ratio distributions in Figs.~\ref{fig:experiments5} and \ref{fig:runD_series} are too strongly bimodal to compare well with the observed smooth continuum of non-resonant ratios.  The bimodality is a consequence of the breakup of the defg and bcd configurations in the TOI-1136 chain.  We conclude that TOI-1136, even sampled over error space, has too specific an orbital architecture to capture the full diversity of chains in nature.  Here we further randomize the initial conditions of our integrations to better reproduce the observed $P_2/P_1$ histogram.

\subsection{A Library of Chains}
\label{subsec:library}

We construct a ``library'' of four-planet and six-planet chains as follows.  Planet masses are randomly drawn from a uniform distribution between $3 M_\oplus$ and $8 M_\oplus$.  For each chain, the innermost planet is assigned a random orbital period between $4$ and $6$ days.  Periods of remaining planets are drawn from a set of resonances: mainly the 3:2 and 2:1, as suggested by the observed young systems \citep{Dai2024}, but also the 6:5, 5:4, 4:3, and 7:5 (proportions to be specified below).  The sequence of resonances in every chain is random, unlike in our TOI-1136-based experiments.

The library comprises 2 types of chains:

\begin{enumerate}
    \item[(i)] Six-planet (``long'') chains (numbering $N_{\rm i} = 256$), each mandated to have three 3:2 pairs, one 2:1 pair, and one pair randomly selected to have a period ratio of 6:5, 5:4, 4:3, or 7:5.
    Type (i) is similar to the TOI-1136-based \texttt{Run D} chains, except that we allow for more resonances in addition to the 2:1, 3:2, and 7:5, and the sequence of resonances from innermost pair to outermost pair is random.
    \item[(ii)] Four-planet (``short'') chains ($N_{\rm ii} = 512$), where relative proportions of resonances are enforced over the ensemble, and not within individual chains, to introduce more randomness.
    We mandate that across $N_{\rm ii} \times 3 = 1536$ pairs, 480 are 3:2, another 480 are 2:1, and the remaining 576 pairs are randomly drawn from 6:5, 5:4, 4:3, and 7:5.
\end{enumerate}

Orbital elements are prepared as follows.  First, values for $\Delta$ are drawn randomly and uniformly from $+0.001$ to $+0.005$, eccentricities are drawn randomly and uniformly from $0$ and $0.03$, inclinations are randomly drawn from a Rayleigh distribution with dispersion $\sigma = 0.5^\circ$, and remaining orbital angles are drawn randomly and uniformly between $0$ and $2\pi$.  To better place the systems in resonance, we damp eccentricities using the \texttt{modify\_orbits\_forces} routine in \texttt{REBOUNDx} for two e-folding times ($2\tau_{\rm e}$).  Then, to ensure that the chains later destabilize, we ramp the eccentricities back up over $3\tau_{\rm e}$, so that their values at the start of the long-term integrations are similar to eccentricities in TOI-1136, between $\sim$$0.01-0.12$.

All chains are required at the start of the integrations to have at least one librating pair, and to have $\Delta > 0$.  Most pairs tighter than 3:2 have $\Delta \sim 0.3$\%.  The 3:2 and 2:1 pairs are prepared with larger initial $\Delta \sim 0.75$--$2$\%, by manually increasing their $\Delta$ values at regular intervals during the eccentricity damping phase.  For 2:1 pairs in long chains, $\Delta \gtrsim 1.2\%$, while for 2:1 pairs in short chains and all 3:2 pairs, $\Delta \gtrsim 0.75\%$.  Such large initial $\Delta$ values for the 3:2 and 2:1 resonances align better with observed $\Delta$ values, and better reproduce the observed fine structures in the period ratio histogram near these resonances after 100 Myr of integration.  In particular, if the initial $\Delta$'s of the 2:1 pairs in long chains are not large enough, disrupted pairs fill in the 2:1 trough at $\Delta < 0$.  Chains that do not satisfy the above requirements are discarded and re-generated.

Thus prepared, the $N = N_{\rm i} + N_{\rm ii} = 768$ chains are integrated for 100 Myr, with mergers treated the same way as in \texttt{Runs D and D1} (3D debris).  The initial and final $P_2/P_1$ histograms are shown in Figure \ref{fig:long_short_chains}.  For both long and short-chain outcomes, non-resonant period ratios are distributed more smoothly than in the TOI-1136-based runs.  The 2:1 trough is still present among the long chains at $t=100$ Myr, with the population falling from a period ratio of 1.9 to 2.0.  The 768 chains constitute our library from which we will draw to fit  observations.

\subsection{\rlr{Runs E and E1}}
\label{subsec:anlys_runE}

We now draw chains from our library whose outcomes at 100 Myr yield a $P_2/P_1$ histogram most closely resembling the observed histogram.  We select $199$ initially six-planet (type i) chains and $95$ initially four-planet (type ii) chains, for a total number of initial pairs equal to that of \texttt{Run D} (1280 pairs).  We judge goodness of fit using a weighted chi-squared
\begin{equation}
  \chi^2
  = \sum_{i \, | \, O_i > 0} w_i \frac{(O_i - M_i)^2}{O_i}
\end{equation}
where $O_i$ and $M_i$ are observed and modeled bin counts, respectively, and $w_i$ is a custom weight \citep{Baker1984}.  Since we are especially interested in reproducing histogram substructure near the 3:2 and 2:1 resonances, we assign $5\times$ higher weights to each of three bins on either side of these resonances as compared to bins elsewhere.  The selection of chains, drawn without replacement, that minimizes $\chi^2$ constitutes \texttt{Run E}.

Figure \ref{fig:random_exp} compares the initial and final period ratio distributions of \texttt{Run E} to observations.  Of all the experiments reported in this work, \texttt{Run E} reproduces the observed period ratio histogram best, particularly with regards to the smoothness of the non-resonant continuum.  Short chains help fill in the gap left by long chains at period ratios just wide of the 3:2 resonance.  The green-colored short-chain contributions to the histogram (Fig.~\ref{fig:random_exp}b) arise from the tightest resonant pairs (7:5, 4:3, 5:4, and 6:5), merging either with each other or with members of the 3:2 to leave behind systems wide of the 3:2.  The 2:1 trough is reproduced.

\begin{figure*}
  \centering
  \includegraphics[width=\linewidth]{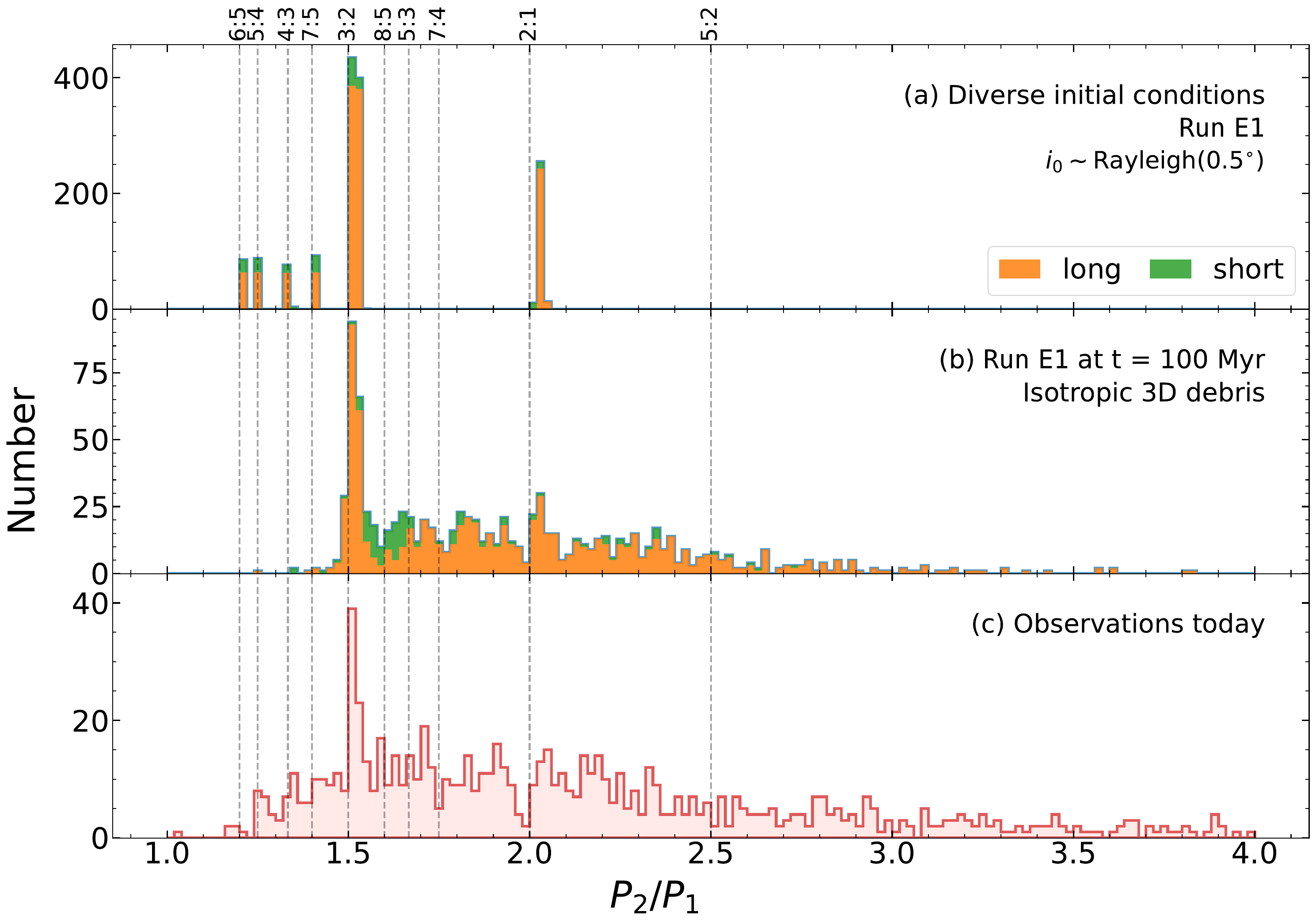}
  \caption{\rlr{Similar to Fig.~\ref{fig:random_exp}, but showing an alternative, simpler selection of chains from our library. \texttt{Run E1} as shown here does not use the $\chi^2$ algorithm underlying \texttt{Run E} of Fig.~\ref{fig:random_exp}, but instead combines all of the type-i long-chain outcomes (unweighted) with all of the type-ii short-chain outcomes that yield at least one planet pair having a period ratio between the 3:2 and 5:3 resonances.  The latter restriction is made in an attempt to fill the $1.54 < P_2/P_1 < 1.67$ gap left by long chains (Fig.~\ref{fig:long_short_chains}, bottom left panel). \texttt{Run E1} does about as well as \texttt{Run E} in reproducing the observations.}
  \label{fig:RunEprime}}
\end{figure*}

\rlr{In many respects the fit to observations in Fig.~\ref{fig:random_exp} seems promising.  To address the concern of excessive cherry-picking by our $\chi^2$ algorithm, we show in Figure \ref{fig:RunEprime} a simpler model that does not use the $\chi^2$ algorithm.  This alternative \texttt{Run E1} is constructed by simply combining all of the type-i long-chain outcomes (unweighted) with all of the type-ii short-chain outcomes that yield at least one planet pair having a period ratio between the 3:2 and 5:3 resonances.  This last restriction is made to try to fill in the $1.54 < P_2/P_1 < 1.67$ gap in the period ratio histogram left by long chains (Fig.~\ref{fig:long_short_chains}, bottom left panel).  Fig.~\ref{fig:RunEprime} shows that the relatively uncurated \texttt{Run E1} reproduces the observations about as well as the curated \texttt{Run E}.  In particular the 2:1 trough is still present in \texttt{Run E1}, an apparently robust consequence of the wide desert of first-order resonances between the 3:2 and 2:1 that disrupted chains find difficult to cross, and 2:1 pairs initialized with large $\Delta \gtrsim 1.2\%$.}

Both \texttt{Run E} and \texttt{Run E1} miss pairs with $P_2/P_1 \lesssim 1.45$ and $P_2/P_1 \gtrsim 3.0$.  The former population could conceivably be generated by stabilizing some tight pairs, perhaps with smaller libration amplitudes at the time of formation, or by isolating tight pairs from other planets by including still shorter chains.  Pairs with the largest period ratios might arise from chains disrupted by planets which for some reason had large eccentricities.

Major and minor merger time distributions for \texttt{Run E} are shown in Figures \ref{fig:merger_time} and \ref{fig:t_minor_from_t_Major}.  Though more major mergers occur later in \texttt{Run E} than in \texttt{Runs B-D}, most still occur too early to compare well with observations (Fig.~\ref{fig:resonance_fraction}).  Outcomes at $t=100$ Myr are comparable, however; the fraction of 1st-order pairs in \texttt{Run E} is $29.5\%$ at $t=100$ Myr, compatible with the observed fraction in this age range.

\section{TTV Phases}
\label{sec:ttv_phases}

\begin{figure*}
  \centering
  \includegraphics[width=0.96\linewidth]{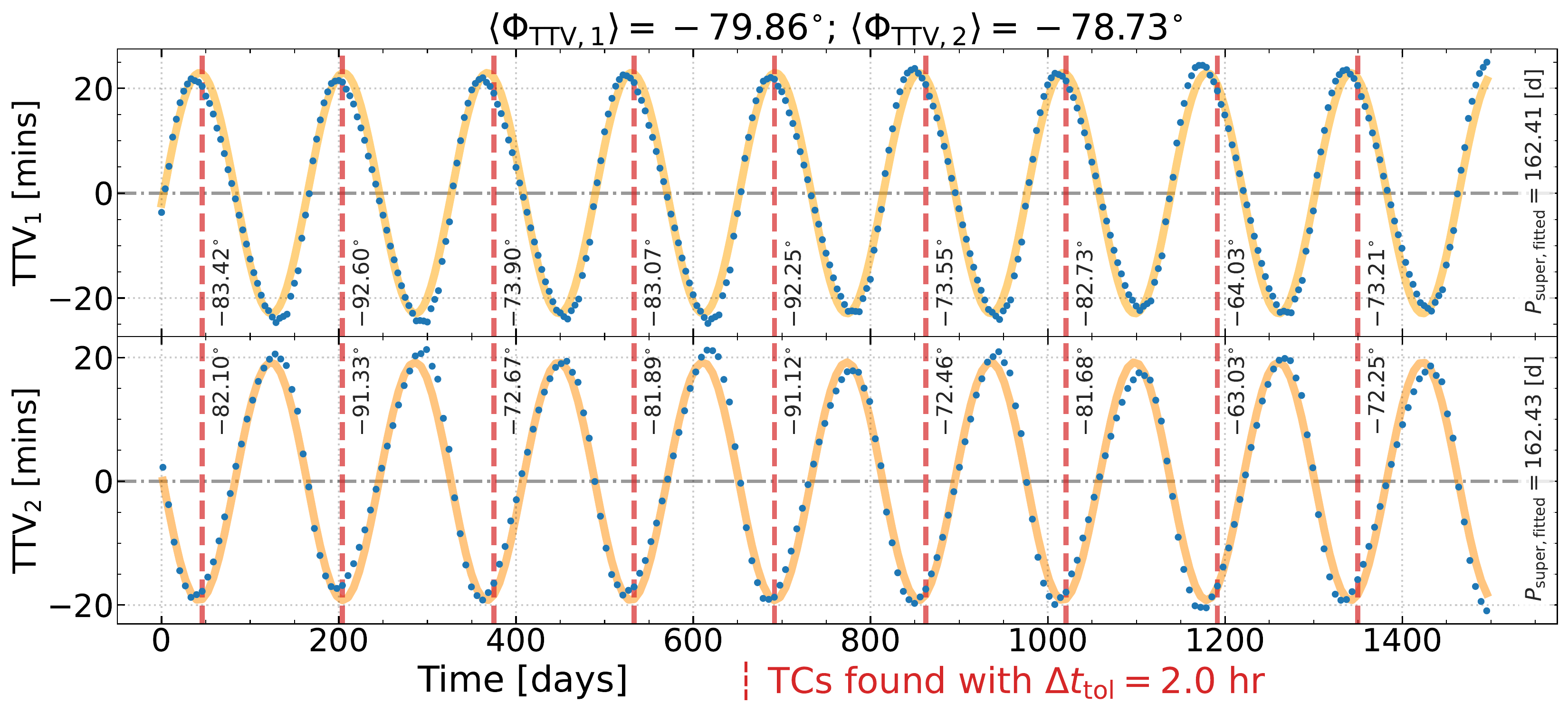}
  \caption{Sample TTV time series from a b:c = 3:2 pair at the end of \texttt{Run D}.  This pair (subscript 1 for the inner member and 2 for the outer member) has a ``well-defined'' TTV phase: transiting conjunctions (marked by vertical dashed red lines) occur consistently before TTV zero-crossings with a time-averaged phase difference of $\langle \Phi_{\rm TTV} \rangle \simeq - 79^\circ$.  Blue points are simulation data and yellow solid curves are sine-wave fits, with the fitted $P_{\rm super}$ labelled on the right.
  \label{fig:good_TTV_phase}}
\end{figure*}

\begin{figure*}
  \centering
  \includegraphics[width=0.95\linewidth]{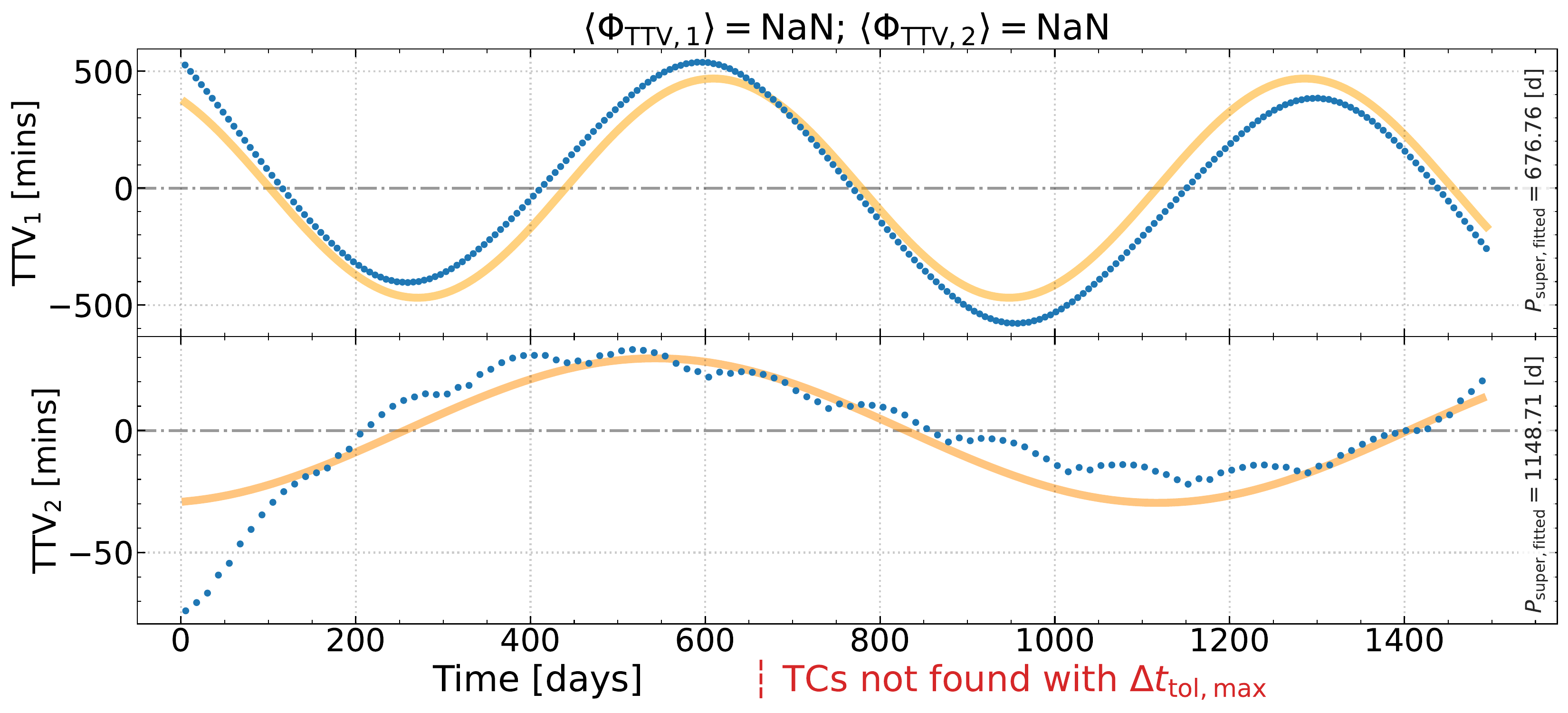}
  \caption{Sample TTV time series for a c:d = 2:1 pair for which a TTV phase is not well-defined.  The sine-wave fits (yellow solid curves) are poor and yield discrepant values for $P_{\rm super}$.  Also, no transiting conjunction is identified, even using the maximum value of $\Delta t_{\rm tol}$.
  \label{fig:bad_TTV_phase}}
\end{figure*}

\begin{figure*}
  \centering
  \includegraphics[width=\linewidth]{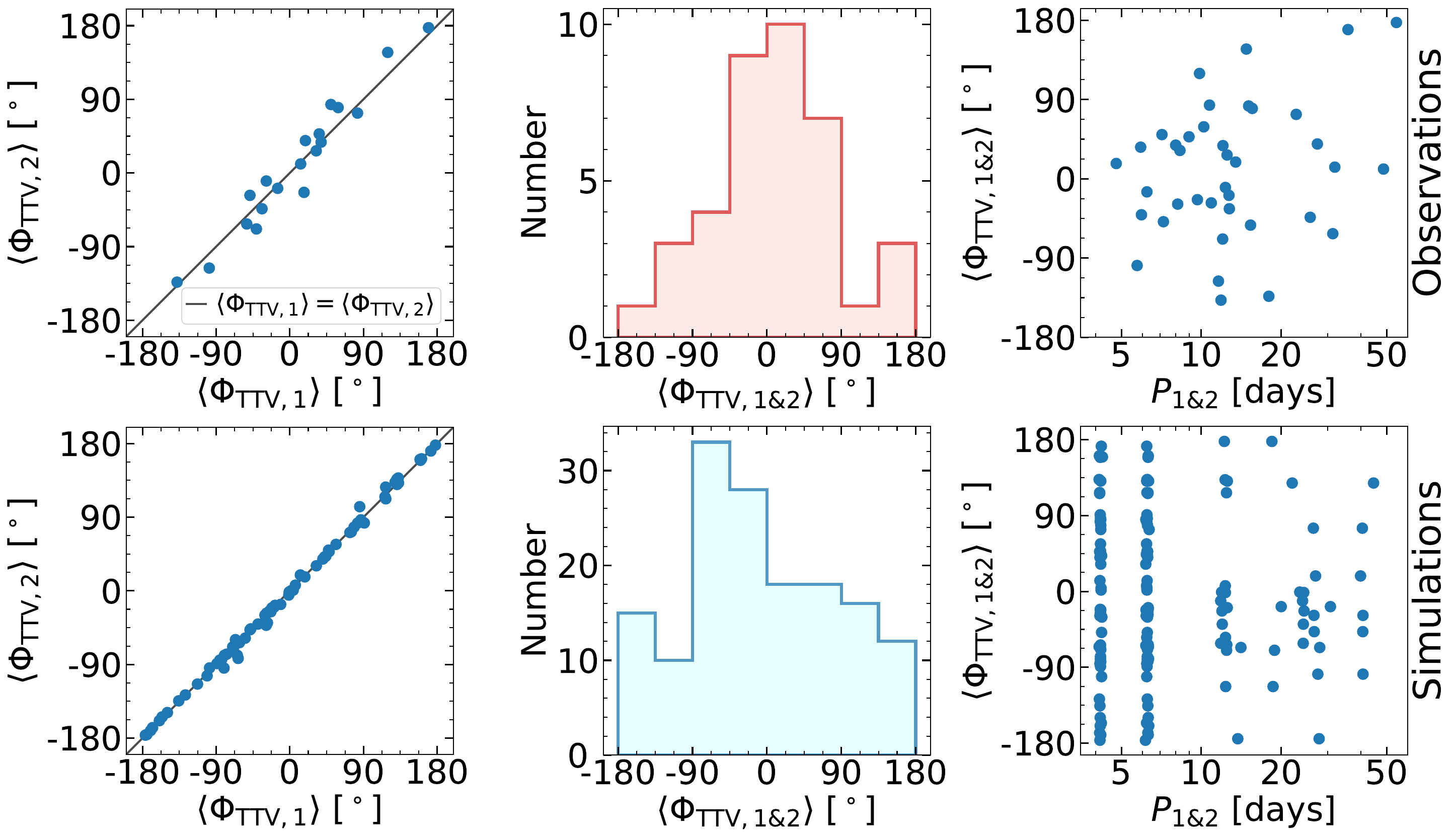}
  \caption{Comparison of time-averaged TTV phases $\langle \Phi_{\rm TTV} \rangle$ between observations (top row) and \texttt{Run D} simulations (bottom row), for planets with $-0.015 \leq \Delta \leq +0.03$ near the 2:1 and 3:2 commensurabilities (the same $\Delta$ criterion for 1st order resonance is used in \citealt{Dai2024}).  Subscript 1 denotes the inner member of the pair, and subscript 2 denotes the outer member.  The TTV phases plotted here are all well-defined, i.e.~consistent between inner and outer members of pairs (left column).  The histograms in the middle column (which plot both $\langle \Phi_{\rm TTV} \rangle_1$ and $\langle \Phi_{\rm TTV} \rangle_2$) show encouraging agreement between observed and simulated TTV phases---a concentration near 0$^\circ$, and broad wings that extend to $\pm$$180^\circ$.  We detect no trend between TTV phase and orbital period, either in the observations or in the simulations (right column).  The simulated resonant pairs all derive from TOI-1136 analogues; most of those with well-defined TTV phases are the innermost surviving b:c = 3:2 pairs.  The list of objects/simulations and their TTV phase curves are available  \href{https://www.rixinli.com/TTVs/MMR-TTV_v3.html}{here} for the top row and \href{https://www.rixinli.com/TTVs/debrisSph-MMR-TTV-v2.html}{here} for the bottom row.
  \label{fig:TTV_Phases}}
\end{figure*}

\begin{figure*}
  \centering
  \includegraphics[width=0.95\linewidth]{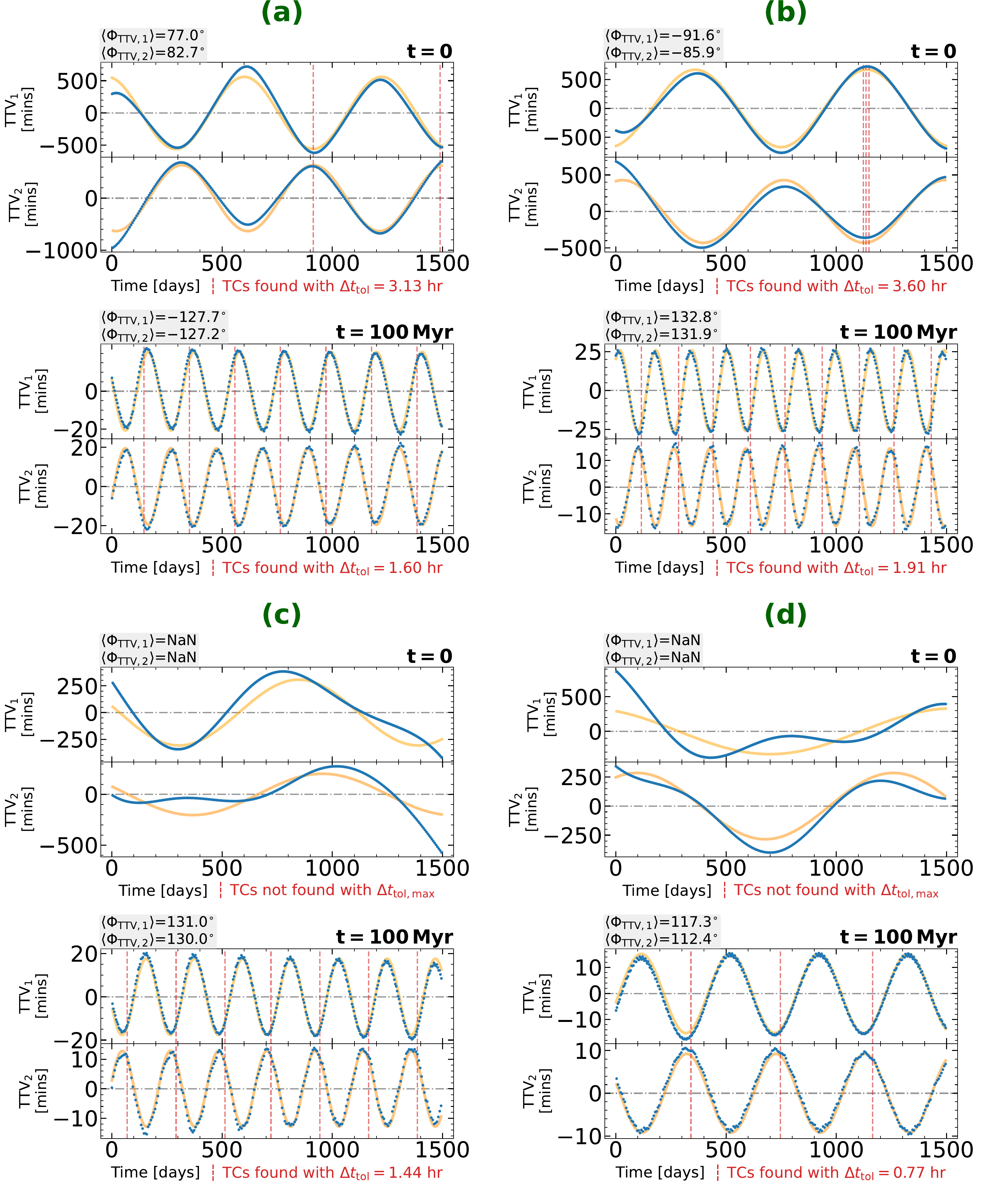}
  \caption{\rlr{Transit timing variations of four 3:2 pairs from \texttt{Run D}, labeled (a)--(d), that survived the full 100 Myr duration of the integrations.  Their TTVs at $t = 100$ Myr (bottom halves of each panel) differ strongly from initial behaviors at $t = 0$ (top halves), a consequence of the wholesale disruption and reorganization of parent chains.  After 100 Myr of mergers, systems contain fewer numbers of planets, and resonant pairs are more isolated from their neighbors, allowing their TTV phases to be better defined; see pairs (c) and (d).  The 4 pairs shown here are drawn from the 75 from \texttt{Run D} having well-defined TTV phases at $t = 100$ Myr; of these 75 pairs, only 7 had well-defined TTV phases at $t = 0$.}
  \label{fig:TTV_phases_i2f}}
\end{figure*}

Transit timing variations (TTVs) provide another diagnostic of resonance.  A resonant/near-resonant pair of planets sufficiently isolated from other resonances from other planets exhibits sinusoidal TTVs for which a phase, $\Phi_{\rm TTV}$, may be evaluated \citep{Lithwick_Xie_Wu_2012}.  The TTV phase is the phase difference between when the two planets transit the star simultaneously (a transiting conjunction, abbreviated as TC) and when the sinusoidal TTV crosses zero (from positive to negative values for the inner planet, and vice versa for the outer; see, e.g., \citealt{Choksi2023}, their figures 2--4).  A pair of planets deeply locked in resonance, with zero libration amplitude---i.e.~having purely forced eccentricities, with no free components---has $\Phi_{\rm TTV} = 0$.  Conversely, a circulating or librating pair is free to have $\Phi_{\rm TTV} \neq 0$.

Observed today, the TTV phases of near-commensurate pairs are distributed broadly from -180$^\circ$ to +180$^\circ$ (\citealt{Lithwick_Xie_Wu_2012, Wu_Lithwick_2013, Hadden_Lithwick_2014}; \citealt{Choksi2023}), indicating that many pairs have substantial non-zero free eccentricities.  This result is not predicted by dissipative, eccentricity-damping processes, including disk-planet interactions and stellar tides, which can otherwise correctly predict the observed peak-trough asymmetries in the $P_2/P_1$ histogram (e.g.~\citealt{Lithwick_Wu_2012}; \citealt{Choksi2023}).  Here we examine to what extent our simulations can resolve this tension.

We sample pairs near the 3:2 and 2:1 resonance for \texttt{Run D} and calculate their phases $\Phi_{\rm TTV}$ as follows.\footnote{The same TTV phase analysis applied to \texttt{Runs B1-B3} yields  too few well-defined $\Phi_{\rm TTV}$ (e.g.~23 out of 1016 for \texttt{Run B1}) to allow for statistical comparison.  A well-defined $\Phi_{\rm TTV}$ requires that a resonant pair  be sufficiently isolated from other planets, and in particular not perturbed by another resonance.  This requirement is largely unmet by the long 6-planet chains in \texttt{Runs B1-B3} which hardly disrupt (Fig.~\ref{fig:runB_series}).}  For each pair, we perform mock transit observations, fitting the average mean-motion over the last 4.1 years of the integration (the duration of the {\it Kepler} transit mission) and subtracting off the mean times of transit to obtain each planet's TTV (the deviation of actual mid-transit time from the mean transit time, vs.~time).  Each TTV signal is fitted to a sine wave using the Python \texttt{lmfit} routine, with amplitude and period determined as part of the fit.  The TTV period, sometimes termed the ``super-period,'' is common to both members of an isolated pair of resonant planets.  In our simulations, sometimes a TTV signal from one member of a pair exhibits two sinusoidal frequencies (due to forcing by other planets), in which case we focus on the frequency component that matches that of the other member of the pair.  Times of TCs are then identified as those times when two planets transit the star within some small time interval $\Delta t_{\rm tol}$ of one another.  We start with a maximum value of $\Delta t_{\rm tol}$ given by the sum of the transit durations, $\Delta t_{\rm tol} = 2R_\odot (1/ v_{\rm K,1} + 1/ v_{\rm K,2})$, where $R_\odot$ is the stellar radius and $v_{{\rm K},i}$ is the Kepler orbital velocity of planet $i$.  Since this choice sometimes yields multiple TCs per super-period when only one TC is possible, we decrease $\Delta t_{\rm tol}$ until we can identify as close to one TC per super-period (within the 4.1-yr analysis duration) as possible.  Pairs which just have a few missing or extra TCs out of an otherwise good set are kept for analysis.  The phase difference between a TC and when the TTV crosses zero (going down for the inner planet, and going up for the outer) is recorded as $\Phi_{\rm TTV}$.

We say that a pair's TTV phase is ``well-defined'' if: (a) the sine-wave fit to each member's TTV time series yields an $R^{2} > 0.9$ according to \texttt{lmfit} (we also double-check that the fit is good-by-eye); (b) the best-fit $P_{\rm super}$ from one member matches that of the other member to within 20\%; and (c) the mean dispersion in $\Phi_{\rm TTV}$ for both members of the pair does not exceed 45 degrees over the 4.1-yr analysis duration (the individual time-averaged $\langle \Phi_{\rm TTV} \rangle$ of one member of the pair may differ from that of the other).  Figure \ref{fig:good_TTV_phase} shows the TTV signals from a b:c pair from \texttt{Run D} where $\Phi_{\rm TTV}$ is well-defined.  By comparison, Figure \ref{fig:bad_TTV_phase} shows an example where the TTV phase is not well-defined---the sine-wave fit to one of the TTV time series is poor, $P_{\rm super}$ determined from one member of the pair differs from $P_{\rm super}$ determined from the other member by almost a factor of 2, and no TC could be found, not even for $\max \Delta t_{\rm col}$.  In \texttt{Run D}, of 163 pairs situated near the 3:2 and 2:1 resonances with $-0.015 \leq \Delta \leq 0.03$---these mostly include unmerged planets b:c, c:d, d:e, and f:g---there are 75 pairs with well-defined TTV phases and 88 pairs without well-defined phases.  Those pairs without well-defined phases are frequently situated adjacent to another near-commensurate pair---the resonances interfere with one another to render their TTV time series non-sinusoidal, as is the case for resonant chains today like TOI-1136 \citepalias{Dai2023} and TRAPPIST-1 \citep{Agol2021}.

In the bottom row of Figure \ref{fig:TTV_Phases}, we show the distribution of well-defined TTV phases from \texttt{Run D}.  These are compared against phases calculated using observed TTVs from \citet{Rowe2015}, as shown in the top row.  The observed TTV phases are computed following the same procedure as described above, with $\max \Delta t_{\rm tol}$ set to the sum of the observed transit durations as given by the March 2024 NASA Exoplanet Archive \citep{cumulative}.  Of the 98 pairs of planets in the Archive with radii $< 4 R_\oplus$ and $-0.015 \leq \Delta \leq 0.03$ near the 3:2 and 2:1 commensurabilities, there are 19 pairs with well-defined TTV phases, a percentage within a factor of 3 of that for \texttt{Run D}.  As is the case with our simulated systems, if an observed near-resonant pair does not have a well-defined TTV phase, it is almost always because its TTV signal is too complex to be fitted with a simple sinusoid, indicating the pair is perturbed by a third body if not more, none of which are necessarily transiting.

Figure \ref{fig:TTV_Phases} shows that our simulations can approximately reproduce the broad distribution of observed TTV phases.  This is as expected, as the dynamical instabilities that lead to mergers and the breaking of resonant chains also perturb surviving resonant pairs.  The resonant survivors are practically fated to have large TTV phases, as the chains start their lives with large libration amplitudes (Figure \ref{fig:theta_32}).  Later major mergers, and minor mergers between resonant planets and debris particles, maintain free eccentricities and TTV phases.  What is perhaps more noteworthy is that our model promises to generate large TTV phases while also producing the prominent 2:1 peak-trough asymmetry in the period ratio histogram (Figure \ref{fig:experiments5}).  In our scenario, there is not just a single dissipative mechanism operating to create the peak-trough asymmetry, as such a mechanism acting alone would also damp TTV phases.  Instead the peak and trough are built and maintained by multiple processes: first, disk-driven migration and eccentricity damping create only peaks at $\Delta > 0$; subsequent dynamical instabilities erode those peaks to form a continuum of period ratios, but one which does not extend to $\Delta < 0$ for the 2:1 resonance; and outside-grazing minor mergers (section \ref{sec:minor}) widen the trough and heighten the peak.  All of this activity cannot help but knock resonant pairs off their fixed points, i.e. excite TTV phases. \rlr{Figure \ref{fig:TTV_phases_i2f} samples four planet pairs from \texttt{Run D} demonstrating that TTV behaviors, including TTV phases, are completely altered from initial conditions over the course of 100 Myr.}

\section{Summary and Discussion}
\label{sec:summary}

Most if not all close-in sub-Neptunes are thought to be born into mean-motion resonant chains, formed by convergent migration of planets in protoplanetary disks (\citealt{Izidoro2017,Izidoro2021,Dai2024}).  These chains are inferred from observations to break on a timescale of $\sim$100 Myr \citep{Dai2024}.  Why the chains break---whether the chains self-destruct from chaotic interactions after the stabilizing effects of the disk are removed \citep[e.g.][]{Izidoro2021}, or whether an external agent like photoevaporative mass loss from planets is needed \citep[e.g.][]{Matsumoto2020}---remain unclear.  A particular challenge is to reproduce the observed destruction timescale of $\sim$100 Myr, which is one if not several orders of magnitude longer than instability timescales reported from various theoretical scenarios and calculations (e.g.~\citealt{Matsumoto2020}; \citealt{Izidoro2021}; \citealt{Goldberg_etal_2022}).

The chains simulated in this paper also broke too quickly, when we initialized them randomly and arbitrarily over various parameter spaces.  When we instead initialized chains using standard disk-migration and eccentricity damping models, we found they did not break at all.

For most of this work we put aside the question of what breaks the chains, and focussed instead on what planetary systems look like after the chains have broken.  We found by successive adjustment of initial conditions and orbital architectures that we could reproduce, after 100 Myr of orbital integrations, the observed statistics of period ratios in multi-planet systems, including the finer structures near mean-motion resonances.  We could also reproduce the behaviors seen in transit timing variations of observed resonant pairs.  Our findings are:

\begin{figure}
    \centering
    \includegraphics[width=\linewidth]{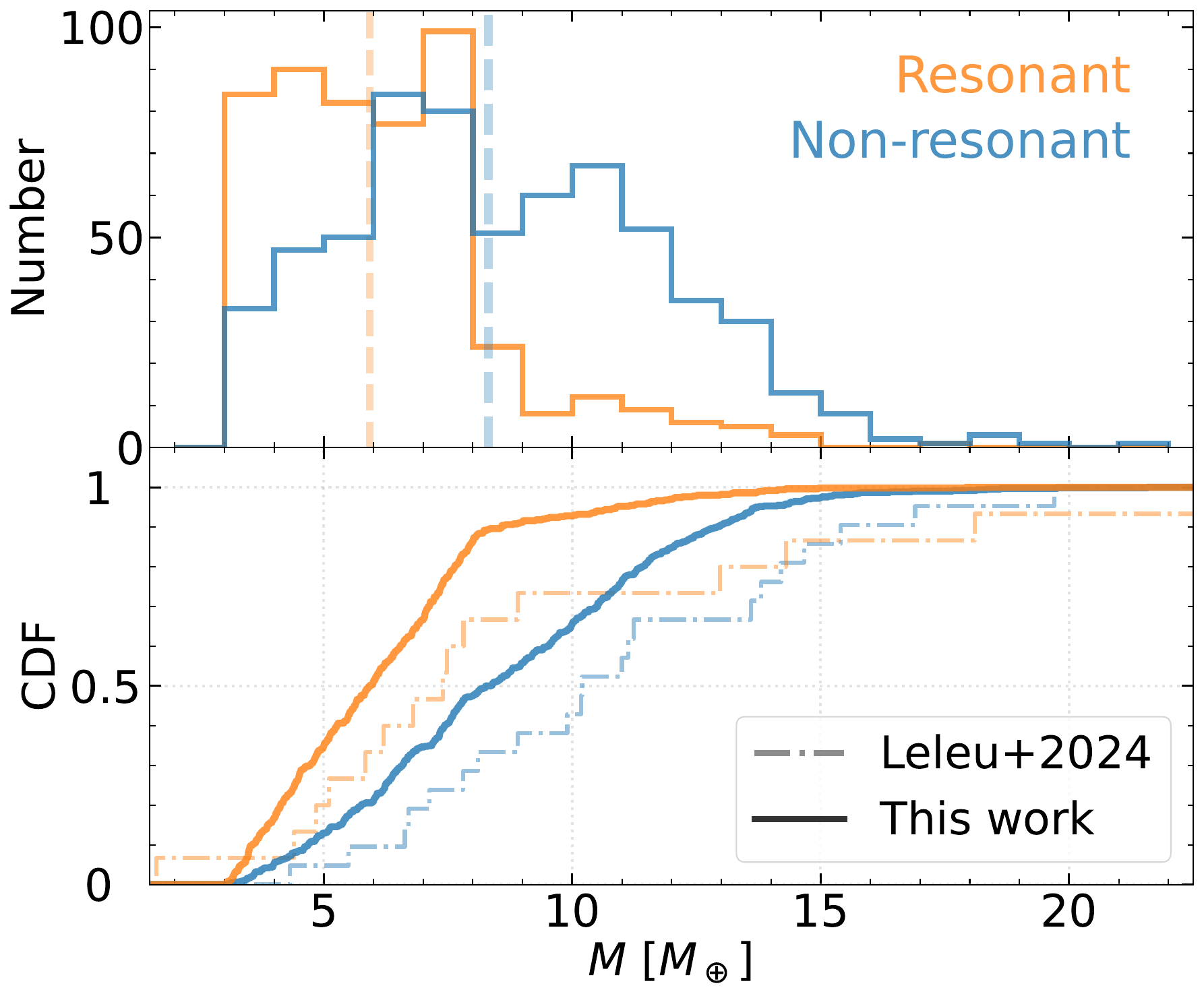}
    \caption{When resonant chains of close-in sub-Neptunes break, non-resonant planets are created from mergers and are more massive than resonant planets which survive mergers.  This is borne out in our simulations, including \texttt{Run E} whose resonant (orange) vs.~non-resonant (blue) planet mass distributions are shown here.  Dashed vertical lines indicate median masses in the top panel histograms, and bottom panel shows cumulative distribution functions (CDFs).  Overplotted in dash-dotted lines are the CDFs of the observed ``controlled sample'' from \citet{Leleu2024}, re-categorized into resonant and non-resonant planets using the \citet{Dai2024} $\Delta$ criteria.  The slope change at $8 M_\oplus$ in our modeled resonant CDF separates the unmerged survivors from merger products that by chance land near a resonance.  A similar slope change is evident in the observed resonant sample.
    \label{fig:final_m_r_dist}}
\end{figure}

\begin{enumerate}
  \item In the histogram of neighboring-planet period ratios $P_2/P_1$ (see Figures \ref{fig:experiments5} and \ref{fig:random_exp}), the observed preference for 3:2 and 2:1 {\it Kepler} planet pairs to be situated just wide of resonance, and the observed deficit of systems just short of the 2:1 resonance (a deficit not observed for the 3:2 resonance), can be reproduced from the following dynamical history:

  Stage i: The resonant peaks in the $P_2/P_1$ histogram form when parent disks drive their planet progeny into chains comprising 3:2 and 2:1 resonances, in addition to an admixture of tighter resonances like the 7:5, 4:3 and 5:4.  Disk-driven migration and eccentricity damping forces the 1st order resonant pairs to librate about their fixed points which lie wide of perfect commensurability; the peaks are centered at $\Delta > 0$, where $\Delta$ is the fractional separation from an $(m+1)$:$m$ period ratio (e.g.~\citealt{Goldreich65}; \citealt{Terquem2019}; \citealt{Choksi2020}).

  Stage ii: Once the disk clears, dynamical instabilities (of unknown origin) over the next 100 Myr cause neighboring planets to merge, with merger products having orbital periods intermediate between those of their progenitors.  Orbital spacings are thus widened.  Merging a member of a 3:2 resonant pair with a planet outside the pair reduces the population in the 3:2 peak and forms a continuum of period ratios at $P_2/P_1 > 3/2$.  An analogous continuum at $P_2/P_1 > 2$ forms from the erosion of the 2:1 peak.  These post-merger continua in period ratio space are smoother when we allow for non-zero mutual inclinations between bodies, and when we include mergers of planets in ``tighter'' (7:5, 4:3, 5:4, 6:5) resonances that are part of shorter (e.g.~4-planet) chains  (Fig.~\ref{fig:random_exp}).

  The continuum arising from the erosion of the 3:2 peak does not extend to indefinitely large $P_2/P_1$ --- the continuum's extent depends on the mass and orbital spacing distribution of the merger progenitors, and the chaotic history of their interactions prior to merging.  If the continuum at $P_2/P_1 > 3/2$ drops off sufficiently before reaching $P_2/P_1 = 2$ --- and note that the distance between the 3:2 and 2:1 resonances is the largest among neighboring 1st order  resonances, with only one 2nd order and two 3rd order resonances located in this space --- then a ``trough'' (population deficit) appears at $\Delta < 0$ for the 2:1 resonance (Figs.~\ref{fig:experiments5} and \ref{fig:random_exp}).  No trough appears shortward of the 3:2 resonance because the distance between the 4:3 and 3:2 resonances is small enough to be bridged by the continuum formed by the erosion of the 4:3 peak.  Likewise the 4:3, 5:4, and tighter resonances do not have troughs.  Encouragingly, in experiments with observed planet architectures other than TOI-1136 (e.g. TOI-178 and TRAPPIST-1; data not shown), and in experiments with diverse random  configurations with 2:1 $\Delta$ values initialized to $> 1\%$ (presumably from Stage i), we also found 2:1 troughs after instability and mergers unfolded.

  Stage iii: Planetesimal debris created from planetary mergers have their eccentricities increased by subsequent gravitational scatterings.  This debris can collide with the orbits of surviving resonant pairs, grazing them from the outside.  When such minor bodies collide with either member of a resonant pair, the pair's orbital spacing is widened, further enhancing the population at $\Delta > 0$ at the expense of $\Delta < 0$ (Fig.~\ref{fig:Delta_bc}).

  \item Free eccentricities in resonant pairs can be excited at every stage of this history, both at the beginning, by whatever process caused the eventual breaking of chains, and also during the ensuing protracted era of dynamical instabilities and giant impacts (major and minor mergers).  Sub-Neptunes perturbed onto eccentric orbits may secularly force neighboring resonant pairs to be similarly eccentric, as shown by \citet{Choksi2023} (in their mechanism, the secularly forced eccentricity from the third-body perturber becomes the resonant pair's free eccentricity).  All of this activity is recorded (and likely repeatedly overwritten) in the transit timing variations (TTVs) of surviving resonant pairs.  If those resonant pairs are sufficiently isolated from other mean-motion resonances with other surviving planets, their TTVs will be simple sinusoids with large phases, as observed.
  
  \item A simple prediction of mergers destroying resonant chains is that non-resonant planets, being merger products, should be more massive than resonant planets, which are mostly unmerged survivors of dynamical instability.  Figure \ref{fig:final_m_r_dist} bears this out for our simulations, and reveals also that a small ($\sim$15\%) fraction of mergers leaves merger products by chance near a resonance.  Observational data from \citet{Leleu2024}  appear to confirm these predictions: mature non-resonant planets are systematically more massive than resonant planets, with resonant planets showing a change in slope in the mass distribution function at $8 M_\oplus$.

\end{enumerate}

\bigskip

\section*{Acknowledgements}

We thank Antoine Petit for an insightful referee's report that led to substantive changes in our work.  R.L. thanks Dong Lai, Chris Ormel, Sharon Xuesong Wang, and Wei Zhu for useful exchanges, and acknowledges support from the Heising-Simons Foundation 51 Pegasi b Fellowship.  E.C. was supported by NSF grant 2205500 and the Simons Foundation Investigator program.  N.C. was supported by a Berkeley Dissertation-Year Fellowship.

\added{\software{Rebound \citep{rebound,reboundmercurius,reboundsa,Lu2024_trace},
                 Matplotlib \citep{Matplotlib},
                 Numpy \& Scipy \citep{Numpy},
                 lmfit \citep{lmfit}.}}

\bibliographystyle{aasjournal}
\bibliography{refs}

\end{document}